\documentclass[journal=nalefd,manuscript=article]{achemso}

\usepackage[version=3]{mhchem} % Formula subscripts using \ce{}

\SectionNumbersOn    % turn off section numbering
\usepackage[capitalize]{cleveref}
\crefname{figure}{Fig.}{Figs.}
\Crefname{figure}{Figure}{Figures}
\crefname{table}{Tab.}{Tabs.}
\Crefname{table}{Table}{Tables}
\crefname{equation}{Eq.}{Eqs.}
\Crefname{equation}{Equation}{Equations}
\crefname{section}{Sec.}{Secs.}
\Crefname{section}{Section}{Sections}

\usepackage{subcaption}
\usepackage{siunitx, soul, xcolor}
\newsavebox{\measurebox}
\usepackage{gensymb}

\usepackage[export]{adjustbox}

\DeclareCaptionFont{capt}{\fontseries{n}\fontfamily{phv}\selectfont}
\usepackage{subfiles} % Best loaded last in the preamble

\author{Shuyu Cheng}
\affiliation{Department of Physics, The Ohio State University, Columbus, Ohio 43210, United States}
\author{M. Nrisimhamurty}
\affiliation{Department of Physics, The Ohio State University, Columbus, Ohio 43210, United States}
\author{Tong Zhou}
\affiliation{Department of Physics, University at Buffalo, Buffalo, New York 14260, United States}
\author{N\'uria Bagu\'es}
\affiliation{Department of Materials Science and Engineering, The Ohio State University, Columbus, Ohio 43210, United States}
\affiliation{Center for Electron Microscopy and Analysis, The Ohio State University, Columbus, Ohio 43210, United States}
\author{Wenyi Zhou}
\affiliation{Department of Physics, The Ohio State University, Columbus, Ohio 43210, United States}
\author{Alexander J. Bishop}
\affiliation{Department of Physics, The Ohio State University, Columbus, Ohio 43210, United States}
\author{Igor Lyalin}
\affiliation{Department of Physics, The Ohio State University, Columbus, Ohio 43210, United States}
\author{Chris Jozwiak}
\affiliation{Advanced Light Source, Lawrence Berkeley National Laboratory, Berkeley, California 94720, United States}
\author{Aaron Bostwick}
\affiliation{Advanced Light Source, Lawrence Berkeley National Laboratory, Berkeley, California 94720, United States}
\author{Eli Rotenberg}
\affiliation{Advanced Light Source, Lawrence Berkeley National Laboratory, Berkeley, California 94720, United States}
\author{David W. McComb}
\email{mccomb.29@osu.edu}
\affiliation{Department of Materials Science and Engineering, The Ohio State University, Columbus, Ohio 43210, United States}
\affiliation{Center for Electron Microscopy and Analysis, The Ohio State University, Columbus, Ohio 43210, United States}
\author{Igor \v{Z}uti\'c}
\email{zigor@buffalo.edu}
\affiliation{Department of Physics, University at Buffalo, Buffalo, New York 14260, United States}
\author{Roland K. Kawakami}
\email{kawakami.15@osu.edu}
\affiliation{Department of Physics, The Ohio State University, Columbus, Ohio 43210, United States}

\title{Epitaxial Kagome Thin Films as a Platform for Topological Flat Bands}


\begin{document}

%%%%%%%%%%%%%%%%%%%%%%%%%%%%%%%%%%%%%%%%%%%%%%%%%%%%%%%%%%%%%%%%%%%%%

  \begin{abstract}
Systems with flat bands are ideal for studying strongly correlated electronic states and related phenomena.
Among them, kagome-structured metals such as CoSn have been recognized as promising candidates due to the proximity between the flat bands and the Fermi level. A key next step will be to realize epitaxial kagome thin films with flat bands to enable tuning of the flat bands across the Fermi level via electrostatic gating or strain.
Here we report the band structures of epitaxial CoSn thin films grown directly on insulating substrates.
Flat bands are observed using synchrotron-based angle-resolved photoemission spectroscopy (ARPES). 
The band structure is consistent with density functional theory (DFT) calculations, and the transport properties are quantitatively explained by the band structure and semiclassical transport theory.
Our work paves the way to realize flat band-induced phenomena through fine-tuning of flat bands in kagome materials.
  \end{abstract}

\section*{}

Strongly correlated electronic systems are one of the focuses of condensed matter physics due to the emergence of interesting many-body ground states.
Materials with dispersionless bands, i.e. flat bands, are ideal systems for studying the physics of strongly correlated electronic states due to the smaller bandwidth $W$ as compared to the Coulomb repulsion $U$.
One noted example is the flat band in twisted bilayer graphene, which is responsible for various correlated phenomena such as tunable superconductivity~\cite{cao2018unconventional, yankowitz2019tuning}, magnetism~\cite{wolf2019electrically} and metal-to-insulator transitions~\cite{lu2019superconductors}.
Another important class of materials exhibiting flat bands are those composed of quasi-two-dimensional (2D) kagome lattices. Examples in this family include CoSn~\cite{kang2020topological, liu2020orbital}, Fe$_3$Sn$_2$~\cite{lin2018flatbands}, Co$_3$Sn$_2$S$_2$~\cite{yin2019negative}, YMn$_6$Sn$_6$~\cite{li2021dirac} and Ni$_3$In~\cite{ye2021flat}.
In 2D kagome lattices, the flat band emerges due to the destructive phase interference of the electronic wave functions within the hexagons of the kagome lattice.
This mechanism generates electronic states confined within these hexagons in real space and appears as non-dispersing bands in momentum space~\cite{li2018realization, kang2020topological, liu2020orbital}.
Theoretically, a 2D kagome lattice generates a perfect flat band in the tight-binding model considering only nearest-neighbor hopping.
In real materials, the actual band structure can deviate from the ideal case due to the existence of additional hopping terms and spin-orbit coupling~\cite{kang2020topological}.

For realizing flat-band-induced phenomena in kagome metals, it is important to fine-tune the flat band position relative to the Fermi level since many physical properties are dominated by states at the Fermi level.
To this end, synthesizing epitaxial thin films of kagome metals provides several strategies for tuning the flat bands.
The highly-controlled growth process using molecular beam epitaxy (MBE) allows one to chemically dope the material, while after the growth, device patterning and voltage gating is another way to tune the flat bands. 
In addition, anisotropic strains can be applied to thin films through either epitaxial growth or mechanical methods.
All these potential advantages provide strong motivation for investigating flat-band-hosting kagome thin films. 
While tunneling spectroscopy has provided indirect evidence for the existence of flat bands in epitaxial FeSn films~\cite{han2021evidence}, direct observation of flat bands in kagome thin films has been missing.

In terms of material selection, CoSn stands out due to the existence of flat bands several hundreds of meV below Fermi level and spread across a large portion of the Brillouin zone (BZ)~\cite{meier2020flat}.
In CoSn bulk crystals, such flat bands have been confirmed by angle-resolved photoemission spectroscopy (ARPES) experiments~\cite{kang2020topological, liu2020orbital, huang2022flat}.
Recent work further established the connection between flat bands in CoSn and the observed large resistance within the kagome plane as compared to perpendicular to the kagome plane~\cite{huang2022flat}.
Regarding the thin film growth, there has been one work reporting epitaxial CoSn thin films on metallic buffer layers by magnetron sputtering~\cite{thapaliya2021high}.
Although the previously reported sputtered CoSn thin films showed physical properties that are consistent with the bulk crystals, the direct evidence of flat bands in CoSn thin films remains elusive \cite{thapaliya2021high}. 
Furthermore, a challenge for epitaxial kagome films is the difficulty of growing continuous films directly onto insulating substrates,~\cite{hong2022synthesis} which are favored for voltage gating and transport studies.

In this paper, we demonstrate epitaxial CoSn thin films grown on insulating substrates as a promising platform to realize flat-band physics.
The growth of (0001)-oriented CoSn thin films on insulating MgO(111) and 4H-SiC(0001) substrates was enabled by a three-step MBE growth recipe.
Using synchrotron-based ARPES, we directly measured the band structure of the CoSn thin films, and revealed multiple flat bands.
At the $\Gamma$ point, spin-orbit coupling (SOC) gaps were observed between one of the flat bands and the quadratic bands, suggesting the nontrivial topology of this flat band.
Using density functional theory (DFT) calculations, we studied the tunability of the flat bands through carrier doping and found that the calculations are consistent with the ARPES experiments.
Finally, we measured the transport properties of CoSn and quantitatively explained the results using the band structure and semiclassical transport theory.

\begin{figure}
    \subfloat[\label{fig:structure}]{
    \includegraphics[width=0.43\textwidth]{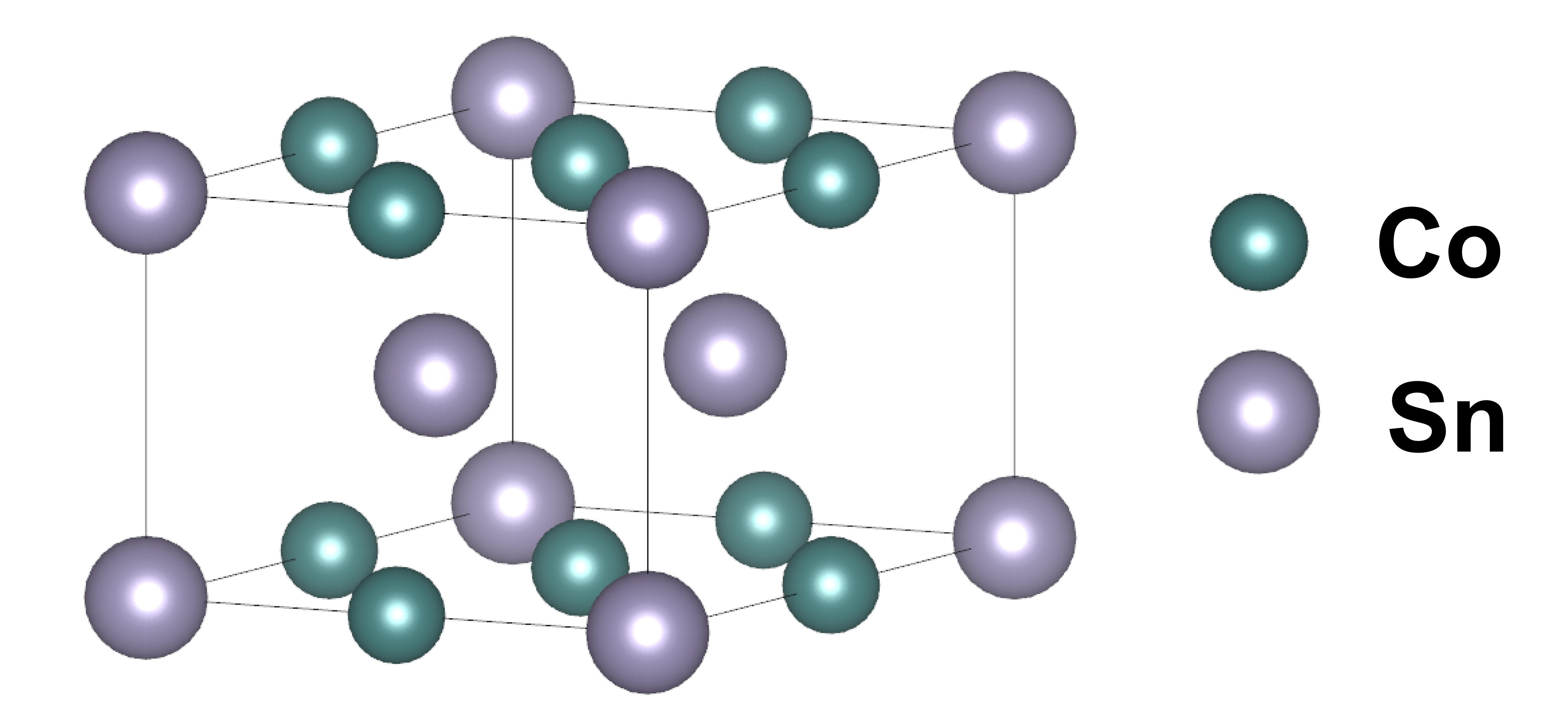}
    }
    \subfloat[\label{fig:RHEED}]{
    \includegraphics[width=0.43\textwidth]{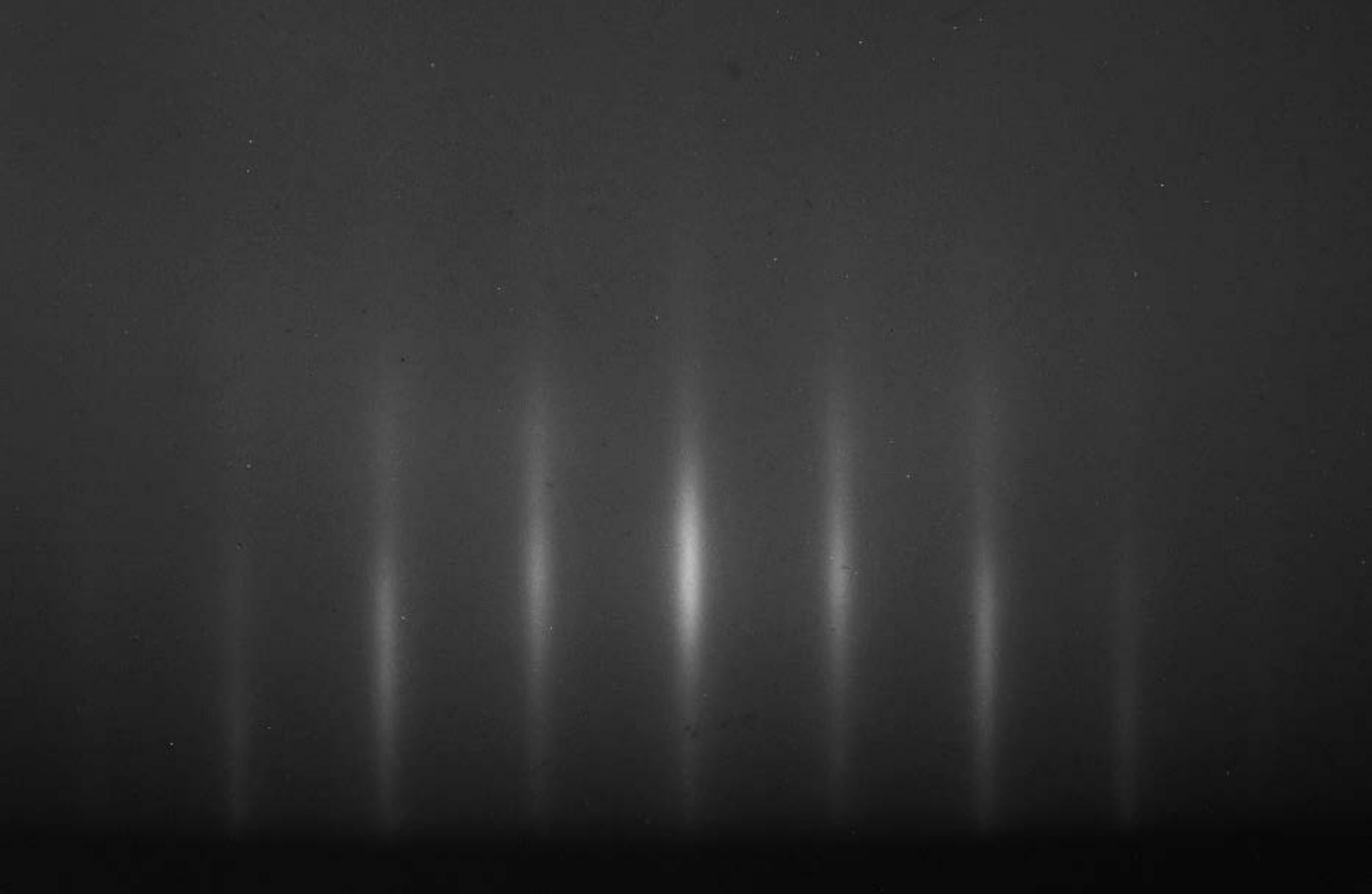}
    }\hfill
    \subfloat[\label{fig:XRD}]{
    \includegraphics[width=0.55\textwidth]{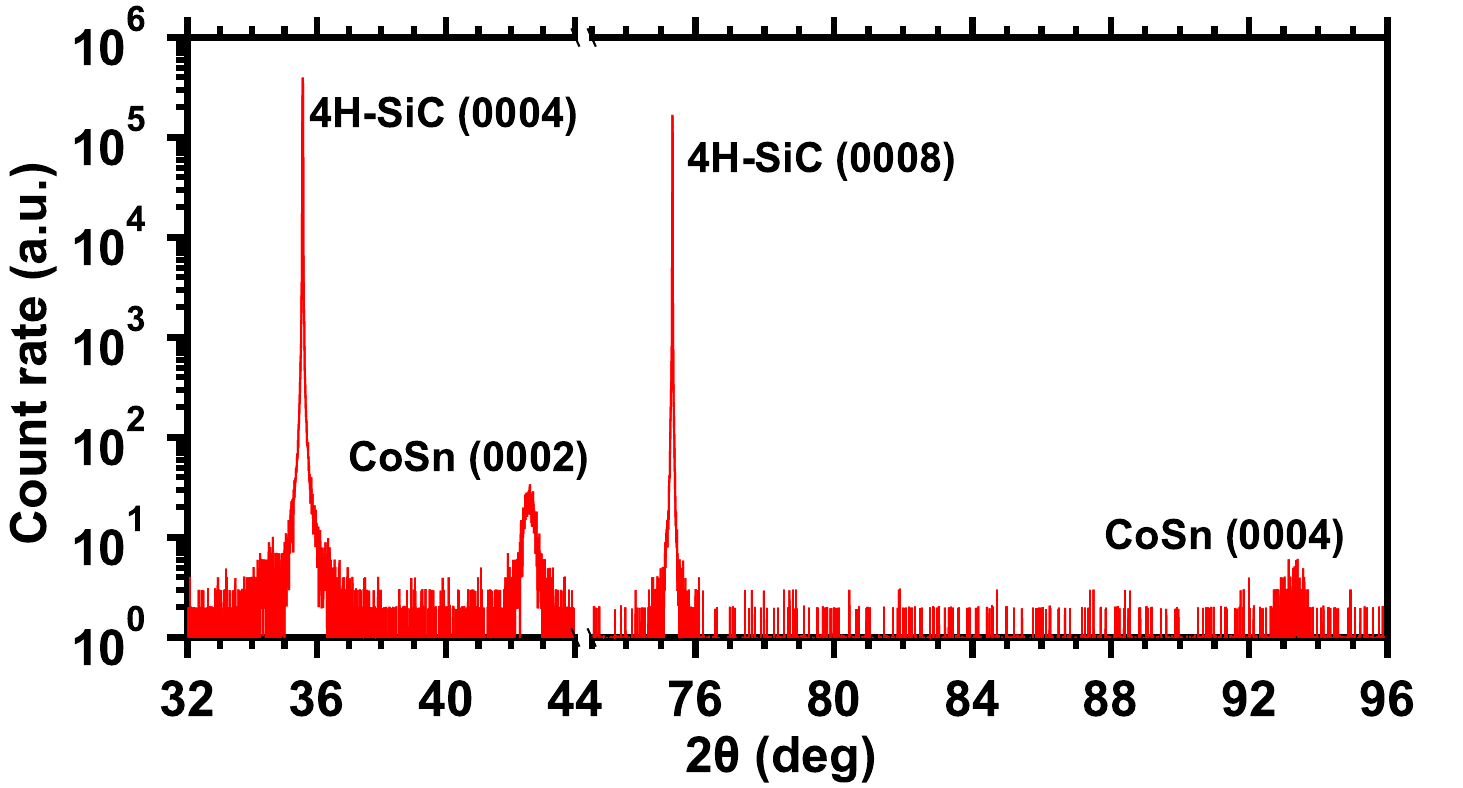}
    }
    \subfloat[\label{fig:TEM}]{
    \includegraphics[width=0.45\textwidth]{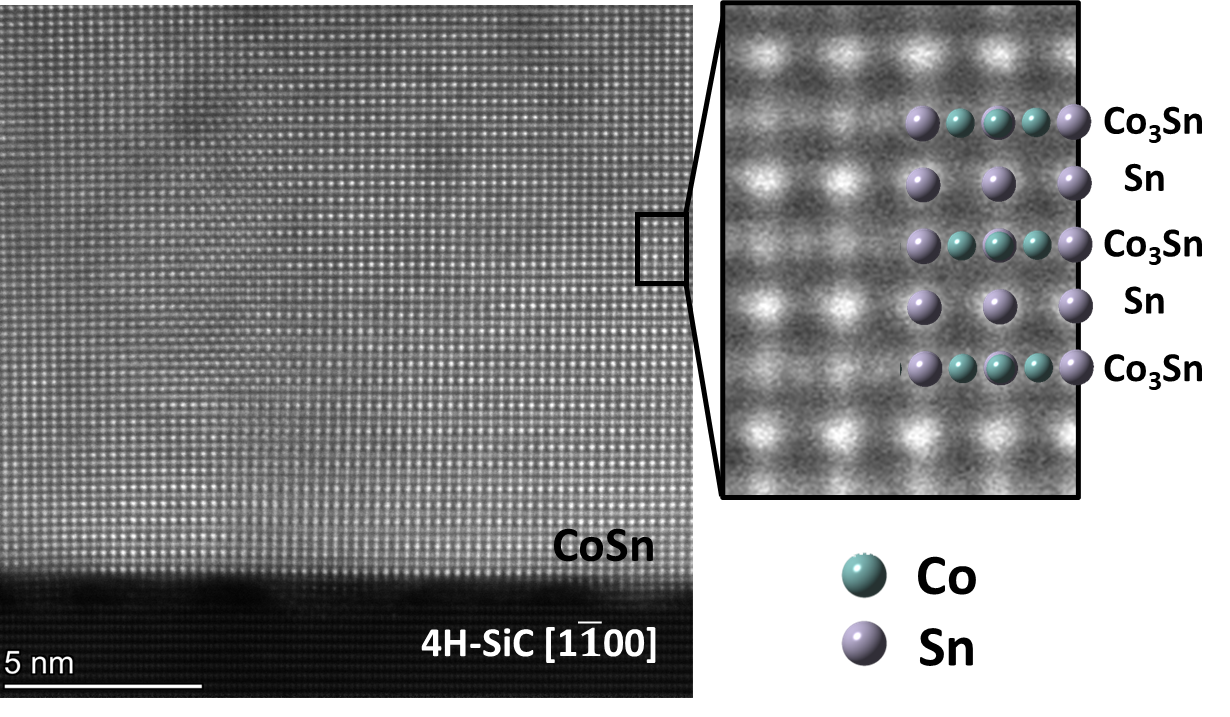}
    }
    \caption{\label{fig:Growth} Sample growth and characterization.
    (a) Crystal structure of CoSn.
    (b) RHEED pattern of a 35\,nm CoSn(0001) thin film grown on 4H-SiC(0001) substrates.
    (c) X-ray diffraction of a 35\,nm CoSn(0001) thin film.
    (d) Atomic-resolution HAADF-STEM image of a CoSn(0001) thin film grown on 4H-SiC(0001) substrate viewed along 4H-SiC[1$\bar{1}$00] direction.
    } 
\end{figure}

The CoSn thin films were grown on MgO(111) or 4H-SiC(0001) substrates by MBE using a three-step recipe. First, a 5\,nm seed layer was deposited at 500\,$^{\circ}$C (470\,$^{\circ}$C) on 4H-SiC(0001) (MgO(111)) substrates, followed by a 15$\sim$20\,nm continuation layer grown at 100\,$^{\circ}$C. The third step is the growth of a terminating layer of 5$\sim$10\,nm CoSn at 300\,$^{\circ}$C.
Details of the growth (and other methods) are provided in the Supporting Information (SI) section S1.

Figure~\ref{fig:RHEED} shows the \textit{in-situ} reflection high-energy electron diffraction (RHEED) pattern of a 35\,nm CoSn thin film grown on 4H-SiC(0001) substrate.
The streaky RHEED pattern indicates epitaxial growth and two-dimensional surfaces with finite terrace width.
The X-ray diffraction (XRD) data of a 35\,nm CoSn thin film grown on 4H-SiC(0001) substrate is shown in Figure~\ref{fig:XRD} (the RHEED and XRD data of CoSn films grown on MgO(111) substrates are provided in the SI section S2).
Besides the substrate peaks at 32.57\,$^\circ$ and 75.34\,$^\circ$, two additional peaks show up at 42.57\,$^\circ$ and 93.34\,$^\circ$, corresponding to CoSn (0002) and (0004) peaks, respectively.
The out-of-plane lattice constant extracted from the XRD scan is 4.254\,\AA, which is in good agreement with the previous studies on bulk crystals~\cite{sales2021tuning} and sputtered thin films~\cite{thapaliya2021high}. 

High-angle annular dark-field (HAADF) scanning
transmission electron microscopy 
(STEM) imaging was performed to examine the crystalline 
characteristics of the CoSn thin film 
(see SI section S1.2 for methods). 
Figure~\ref{fig:TEM} shows an atomic-resolution HAADF-STEM image of a 35\,nm CoSn(0001) thin film on 4H-SiC(0001) viewed along 4H-SiC[1$\bar{1}$00] axis.
The HAADF-STEM image reveals the alternating stacking sequence of one Co$_3$Sn kagome layer and one Sn$_2$ honeycomb layer, which is expected for CoSn~(see Figure~\ref{fig:structure}). 
The brightness of atomic columns in the HAADF-STEM image is approximately proportional to the square of the atomic number (Z), consequently, Sn (Z=50) atom columns appear as brighter columns while a mixture of CoSn atom columns and Co (Z=27) atom columns appear dimmer. 
Although the sequence of the alternating stacking of one Co$_3$Sn layer and one Sn$_2$ layer is predominant across the film, in some regions closer to the interface, the sequence is alerted by the addition of extra Co$_3$Sn layers (See SI section S3). 

\begin{figure}[htbp]
    \subfloat[\label{fig:FB_97eVLH}]{
    \includegraphics[width=0.51\textwidth]{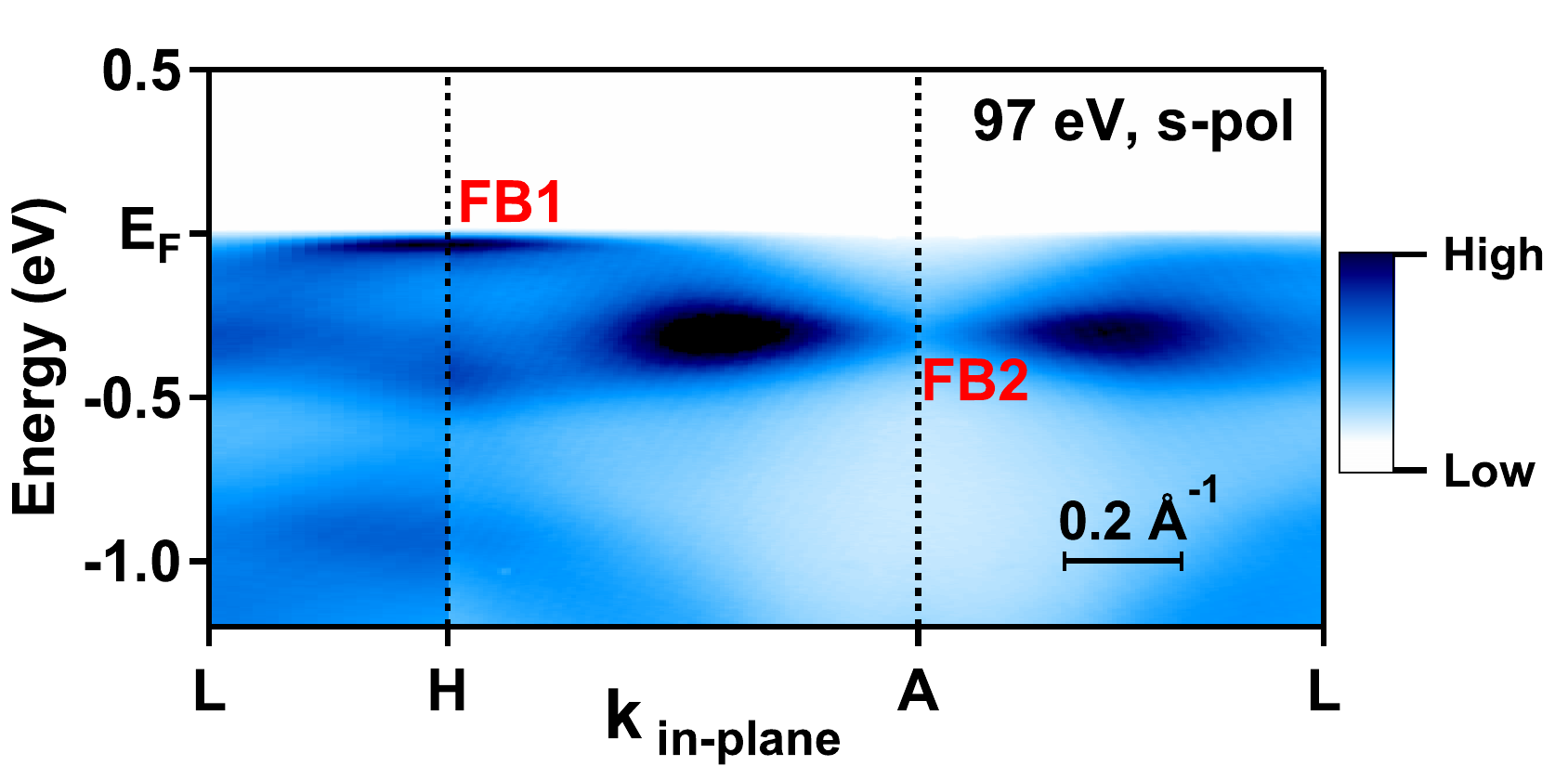}
    }
    \subfloat[\label{fig:FB_128eVLH}]{
       \includegraphics[width=0.462\textwidth]{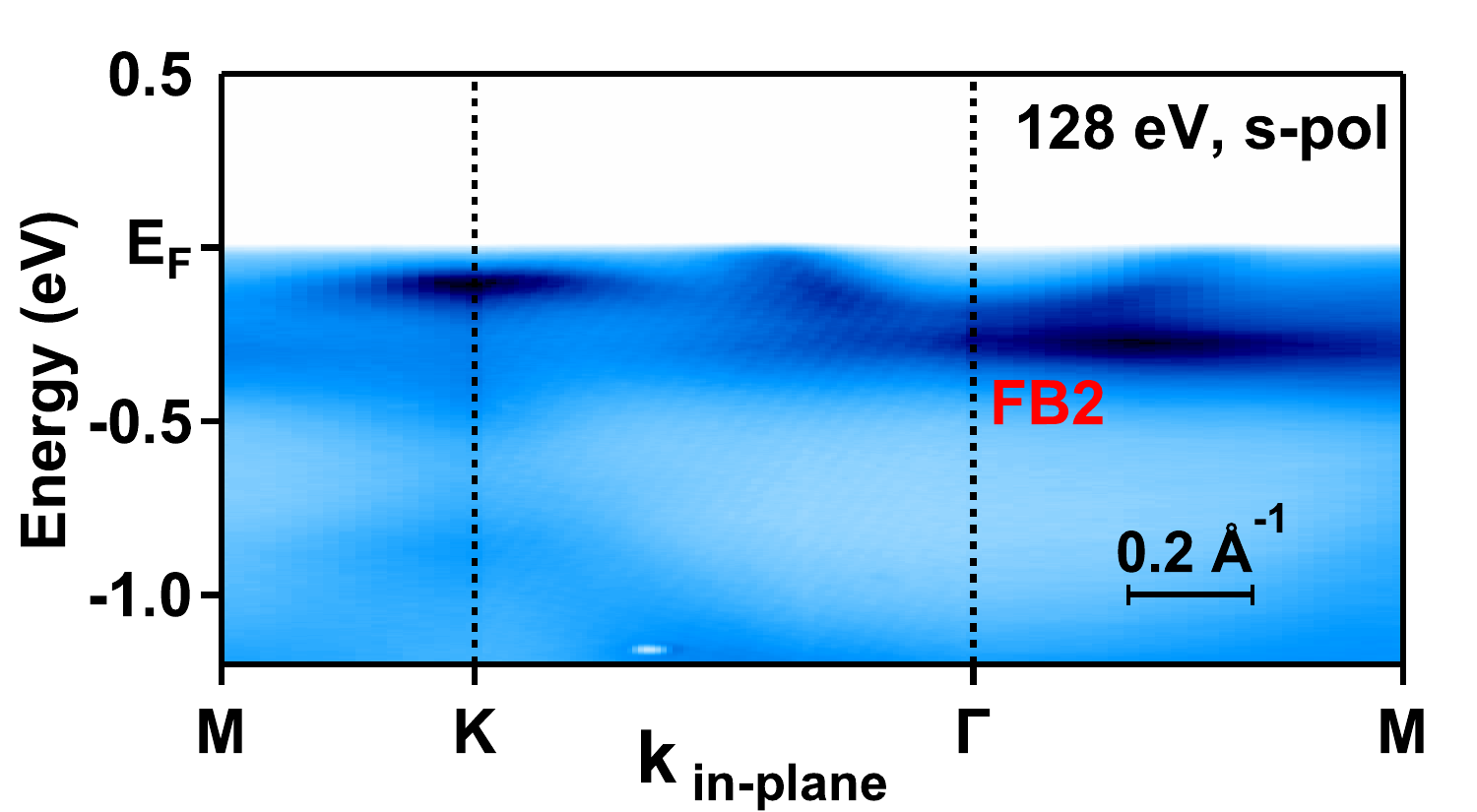}
    }\hfill
    \begin{minipage}{.3\textwidth}
    \subfloat[\label{fig:FermiSurface_128eVLH}]{
       \includegraphics[width=\textwidth]{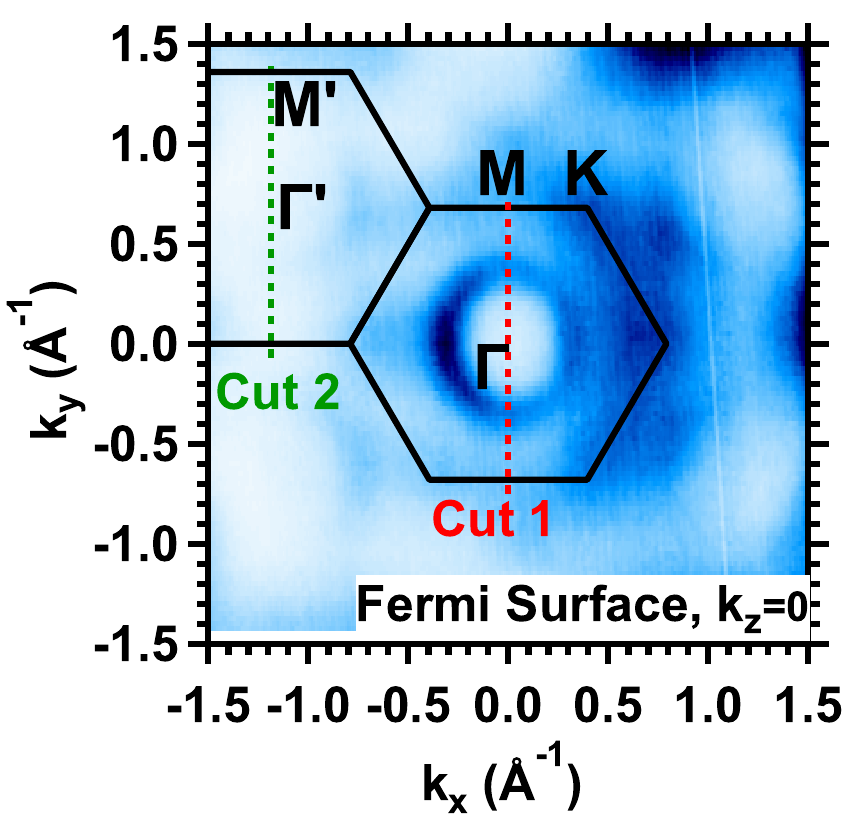}
      }    
   \end{minipage}
   \begin{minipage}{.65\textwidth}
   \subfloat[\label{fig:MGM_1stBZ}]{
       \includegraphics[width=0.5\textwidth]{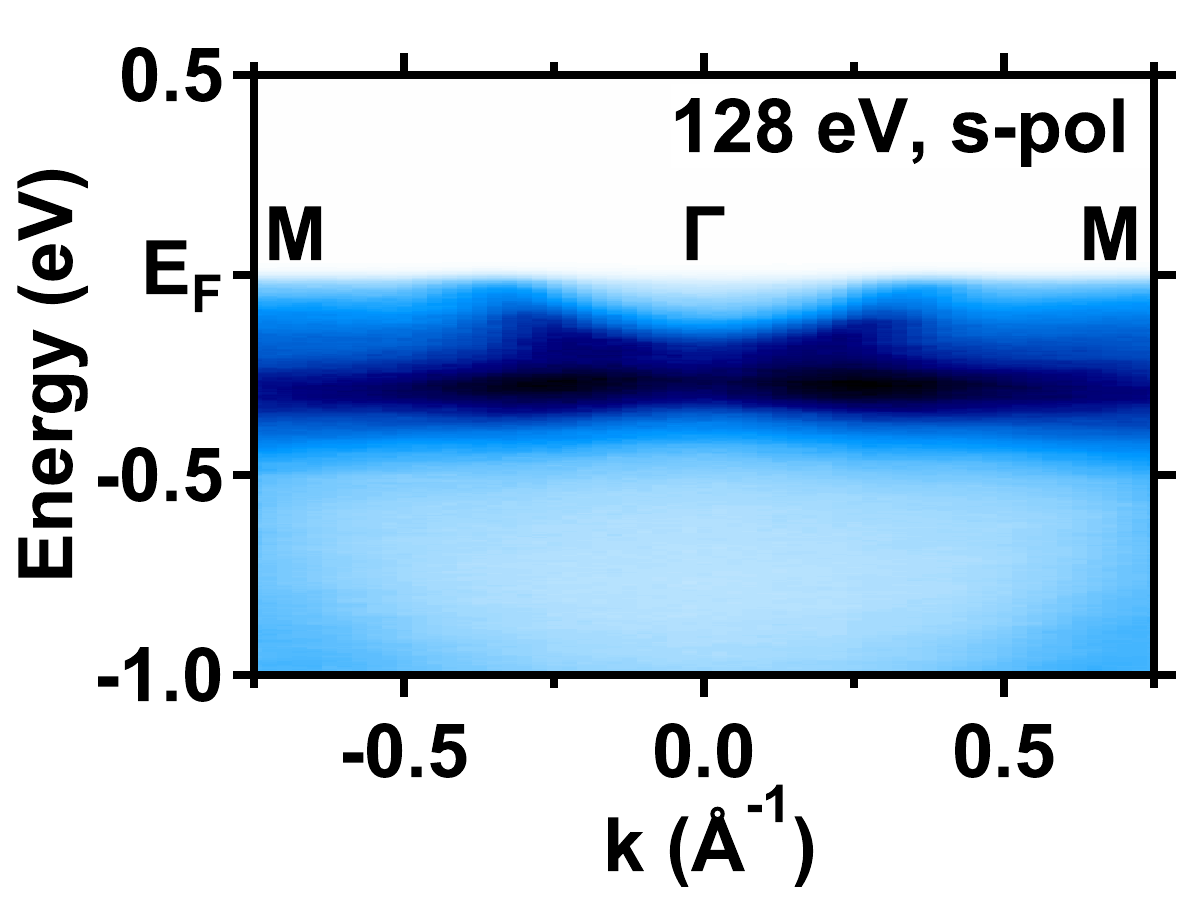}
      }  
    \subfloat[\label{fig:EDC_Stack_MGM}]{
       \includegraphics[width=0.48\textwidth]{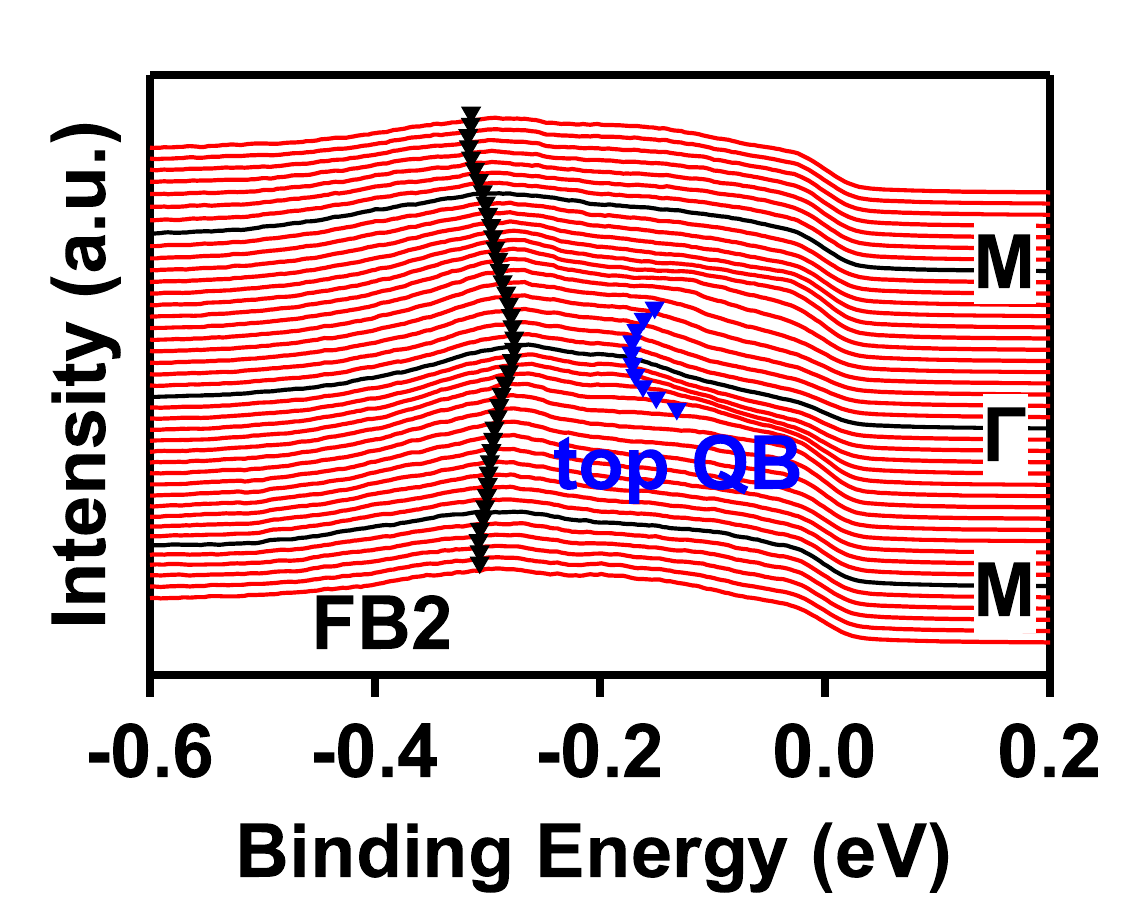}
    }\hfill
    \subfloat[\label{fig:MGM_2ndBZ}]{
       \includegraphics[width=0.5\textwidth]{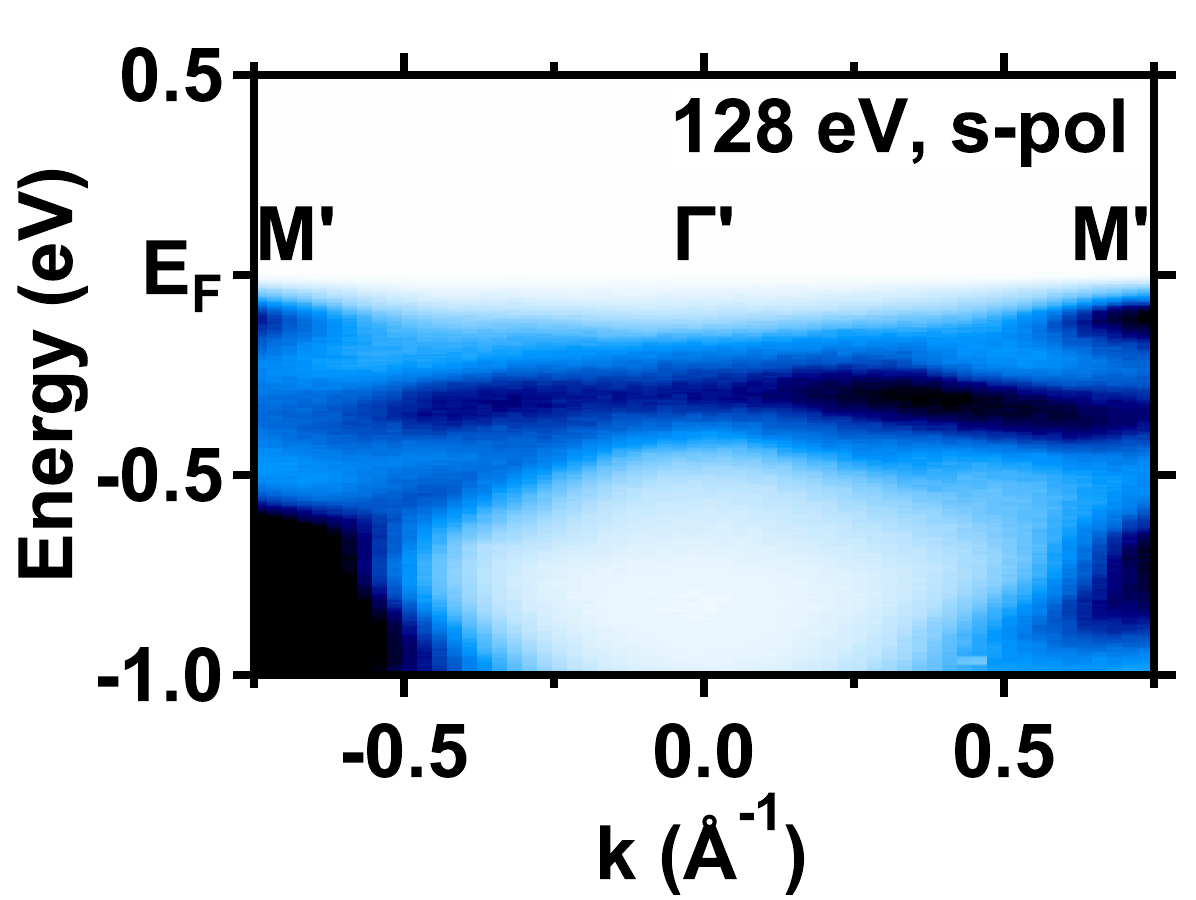}
    } 
    \subfloat[\label{fig:EDC_Stack_MG1M}]{
       \includegraphics[width=0.48\textwidth]{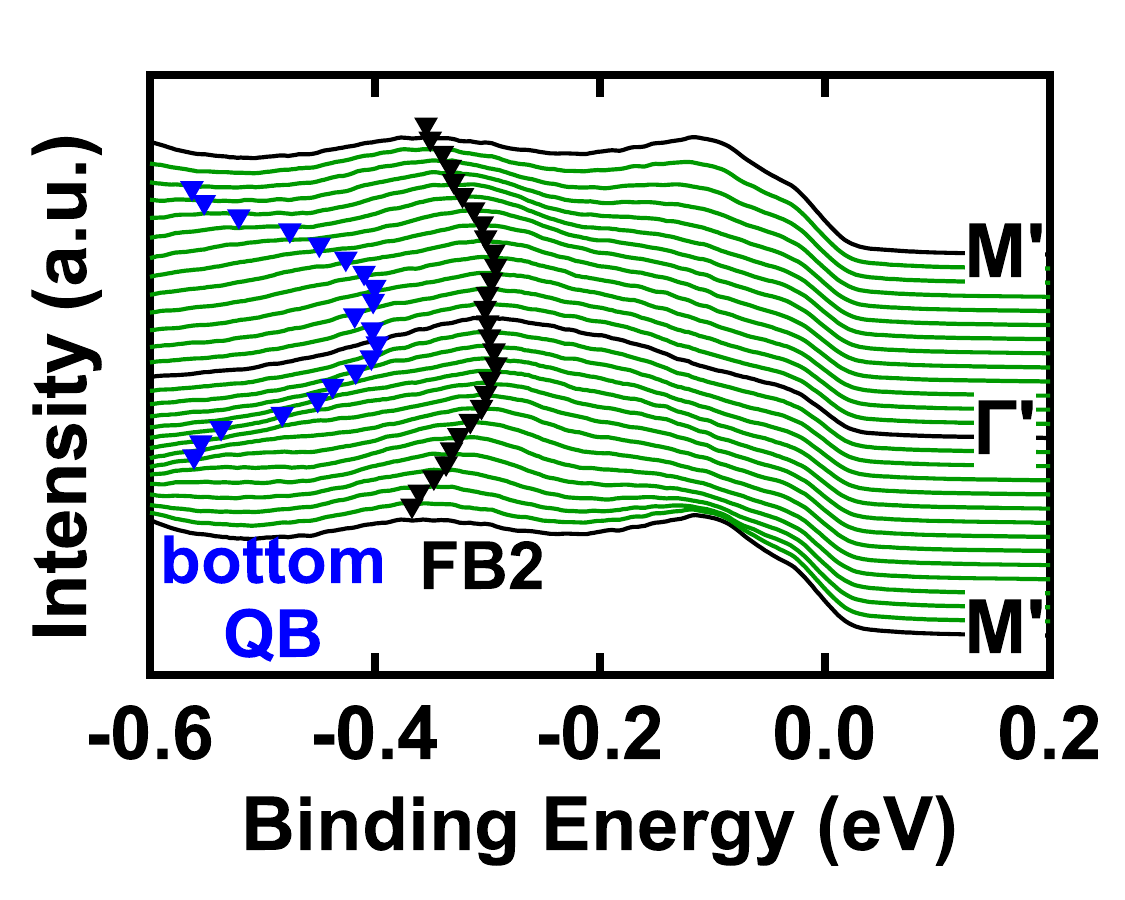}
    }
    \end{minipage}
    \caption{\label{fig:FlatBands} Observation of flat bands in a 25\,nm CoSn thin film on 4H-SiC(0001) substrate.
    (a) ARPES spectra in the $k_z = \pi$ (mod $2\pi$) plane, measured with 97\,eV s-polarized photons.
    (b) ARPES spectra in the $k_z = 0$ (mod $2\pi$) plane, measured with 128\,eV s-polarized photons.
    (c) Fermi surface of $k_z = 0$ (mod $2\pi$) plane. The black hexagons represent BZs. 
    (d) ARPES spectra along the M-$\Gamma$-M direction (``Cut 1" in (c)) in the 1st BZ.
    (e) EDCs of (d). The black and blue delta symbols mark the position of peaks from FB2 and the top QB, respectively.
    (f) ARPES spectrum along the M'-$\Gamma$'-M' direction (``Cut 2" in (c)) in the 2nd BZ.
    (g) EDCs of (f). The black and blue delta symbols mark the position of peaks from FB2 and the bottom QB, respectively.} 
\end{figure}

Having synthesized the epitaxial CoSn thin films, the questions one would raise is whether or not our thin films have flat bands, and whether or not the flat bands are topologically nontrivial if they exist.
The most straightforward method to answer these questions is to directly measure the band structures using ARPES.
In this study, we utilized the synchrotron-based ARPES to map the band structures of a 25\,nm CoSn thin films grown on MgO(111) and 4H-SiC(0001) substrates (see SI section S1.3 for methods).
In the following discussion, we will mainly focus on the band structures of CoSn on 4H-SiC(0001) substrates, while the band structures of CoSn on MgO(111) substrate are presented in SI section S4.1.
Figure~\ref{fig:FB_97eVLH} shows the band structures measured using s-polarized (s-pol) photons with 97\,eV energy, which corresponds to $k_z=\pi$ (mod $2\pi$) plane in the momentum space (see SI section S4.2 for k$_z$ dependence of ARPES spectrum, and SI section S4.3 for ARPES spectra taken with p-polarized light).
In this plane, two dispersionless bands can be observed - one band centers around the H point (labeled as ``FB1" in Figure~\ref{fig:FB_97eVLH}) close to the Fermi level, while the other band (``FB2") resides deeper below the Fermi level and spreads over almost the entire BZ.
Lorentzian fitting of energy distribution curves (EDCs) around the H point gives a binding energy of -0.04\,eV for FB1, with effective mass $m^{*}$=16.7\,$m_{0}$ along the H-L direction, where $m_0$ is the mass of free electrons.
Meanwhile, fitting EDCs at the A and L points yields -0.31\,eV and -0.32\,eV binding energies, respectively (see SI section S4.4 for details).

It is noteworthy that FB1 is reported to be accountable for the anomalous anisotropic transport properties and orbital magnetic moment in bulk CoSn, due to its proximity to the Fermi level\cite{huang2022flat}.
Another work demonstrates that tuning FB1 to the Fermi level through chemical doping induces antiferromagnetic ordering at low temperatures\cite{sales2022chemical}.
In our work, FB1 lies just below the Fermi level at the H points, which is promising for tuning it across the Fermi level in the near future.

We next measured the band structures of CoSn thin films in the $k_z=0$ (mod $2\pi$) plane by changing the photon energy to 128\,eV.
Different from the $k_z=\pi$ (mod $2\pi$) plane in which two sets of flat bands can be observed, in the $k_z=0$ (mod $2\pi$) plane only the FB2 can be seen $\sim$0.3\,eV below the Fermi level.
The fitting results of the EDCs at multiple high-symmetry points across the 1st BZ suggest that the variation of FB2 binding energy is within $\pm$0.03\,eV, signifying the dispersionless nature of FB2 (See SI section S4.4). 

An important question yet to be answered is whether the flat bands are topologically non-trivial for the CoSn thin films.
Theoretically, the spin-orbit coupling (SOC) opens a gap between the flat band and quadratic band (QB) at the band touching point, making the flat band topologically non-trivial~\cite{sun2011nearly, guo2009topological, bolens2019topological}. 
To verify this, we analyzed the spectrum around the $\Gamma$ point, where FB2 touches the QB.
The spectrum along the M-$\Gamma$-M direction (``Cut 1" in Figure~\ref{fig:FermiSurface_128eVLH}) in the 1st BZ clearly captures FB2 and the top QB, as shown in Figure~\ref{fig:MGM_1stBZ}.
Lorentzian fitting of representative EDCs along this direction yields a SOC gap of 0.12\,eV between FB2 and the top QB (Figure~\ref{fig:EDC_Stack_MGM}).
Meanwhile, the spectrum intensity of the bottom QB is maximized in the 2nd BZ along M'-$\Gamma$'-M' direction (``Cut 2" in Figure~\ref{fig:FermiSurface_128eVLH}), as shown in Figure~\ref{fig:MGM_2ndBZ}.
A similar analysis yields a 0.09\,eV SOC gap between FB2 and the bottom QB  (see Figure~\ref{fig:EDC_Stack_MG1M}).
Compared to the reported 0.04 to 0.08\,eV SOC gap reported in CoSn bulk crystals~\cite{kang2020topological, liu2020orbital}, the SOC gap for the thin film is slightly larger.

\begin{figure}[h]
    \subfloat[\label{fig:DFT_prisrtine}]{
    \includegraphics[width=0.49\textwidth]{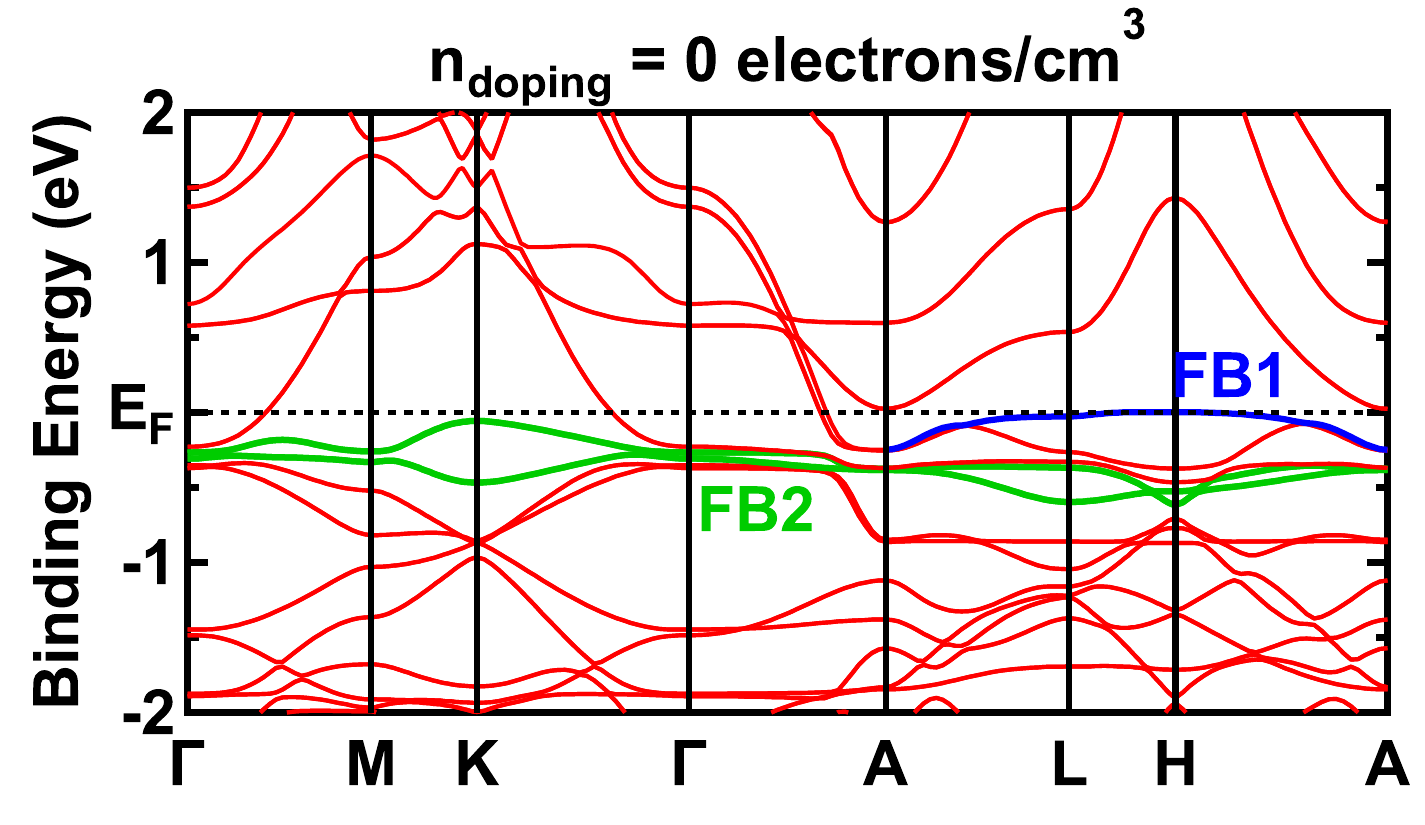}
    }
    \subfloat[\label{fig:DFT_doped}]{
    \includegraphics[width=0.49\textwidth]{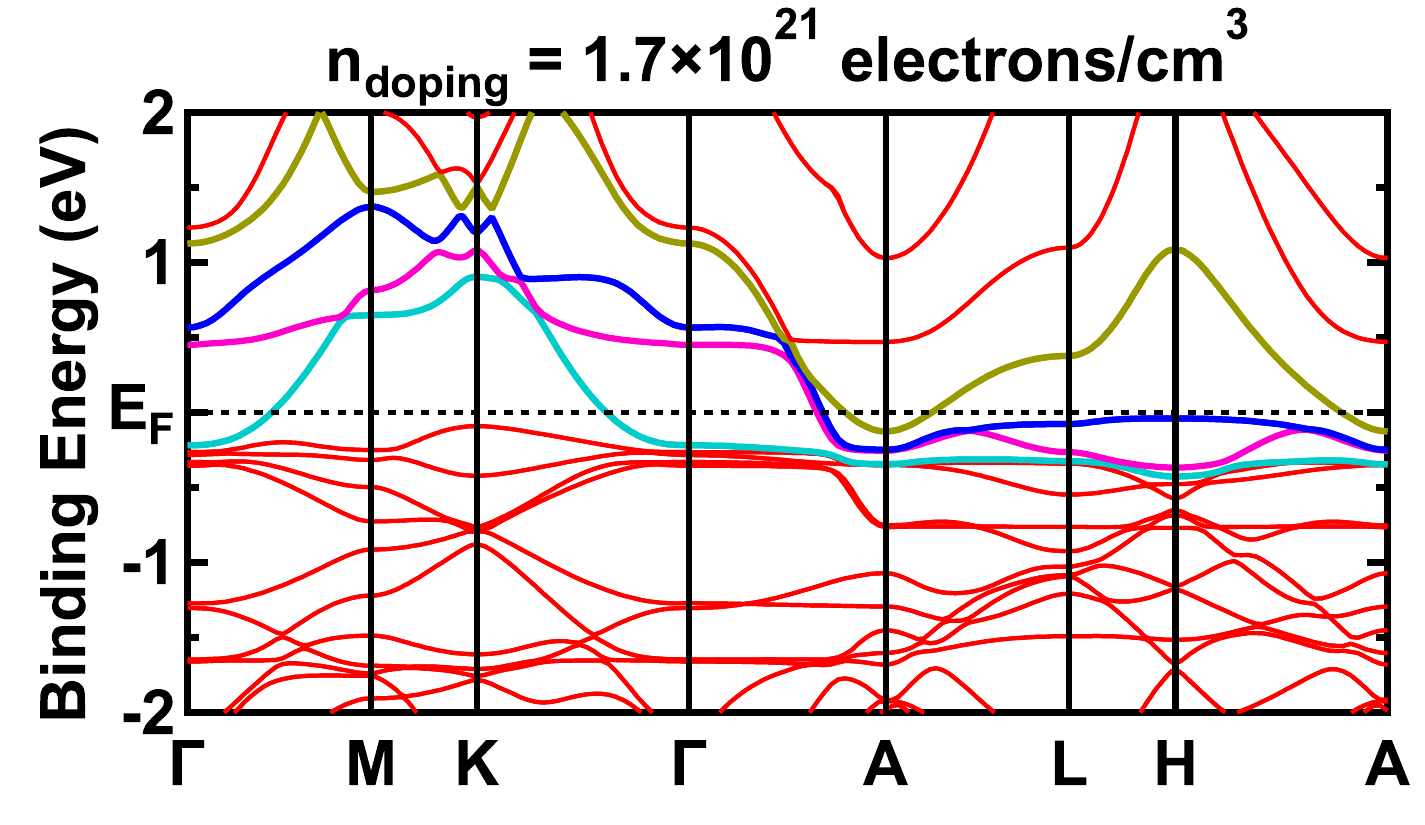}
    }\hfill
    \subfloat[\label{fig:DFT_energy_shift}]{
       \includegraphics[width=0.22\textwidth]{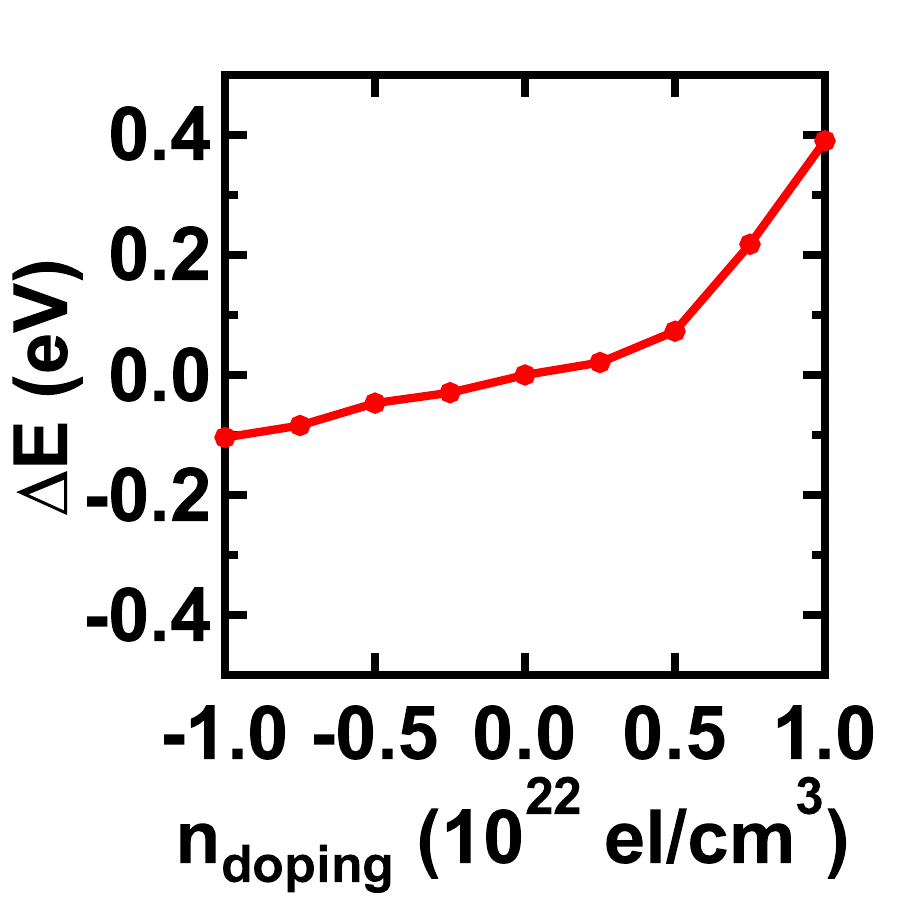}
    }    
    \subfloat[\label{fig:DFT_stack_97eVLH}]{
    \includegraphics[width=0.38\textwidth]{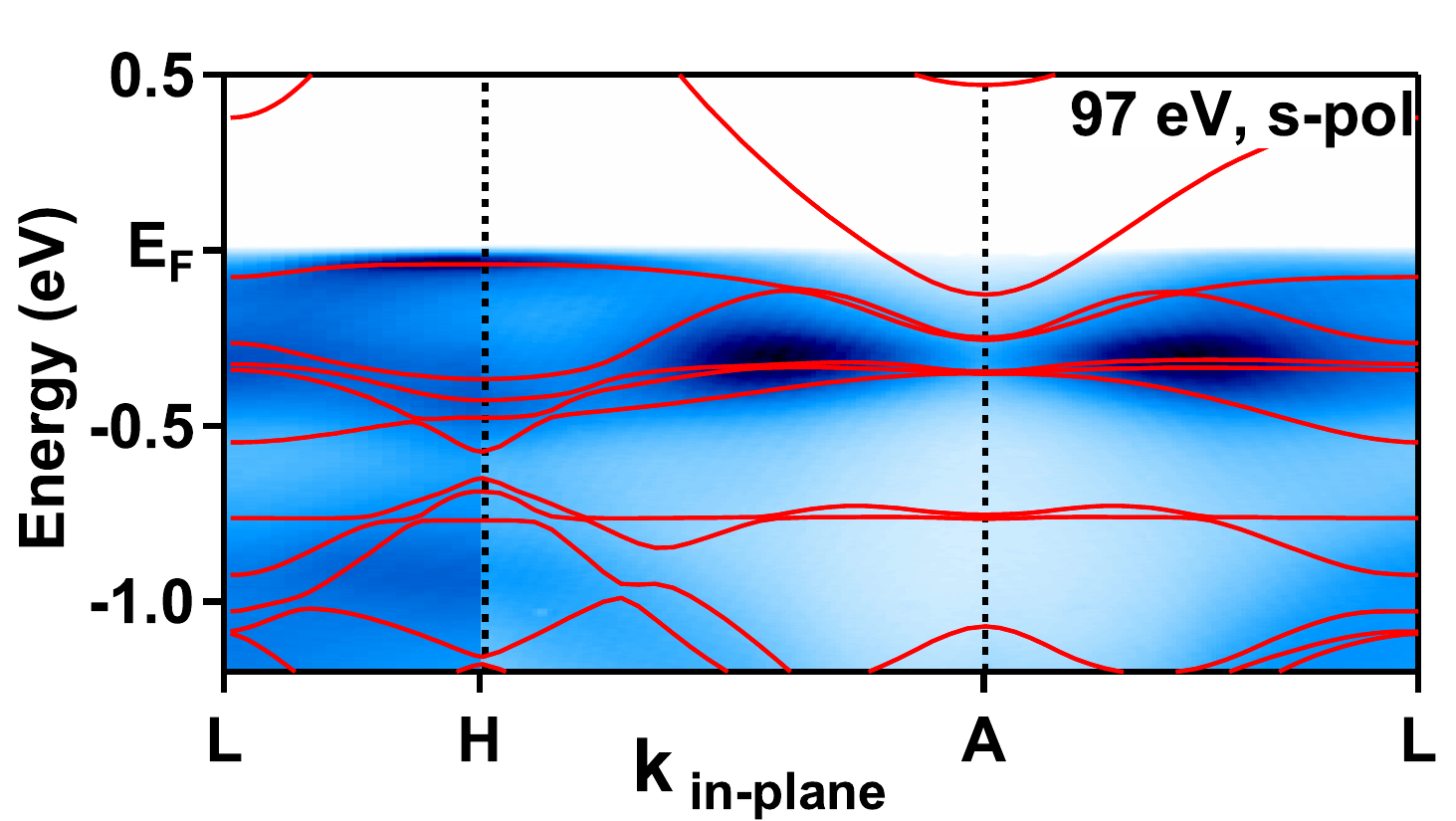}
    }
    \subfloat[\label{fig:DFT_stack_128eVLH}]{
       \includegraphics[width=0.38\textwidth]{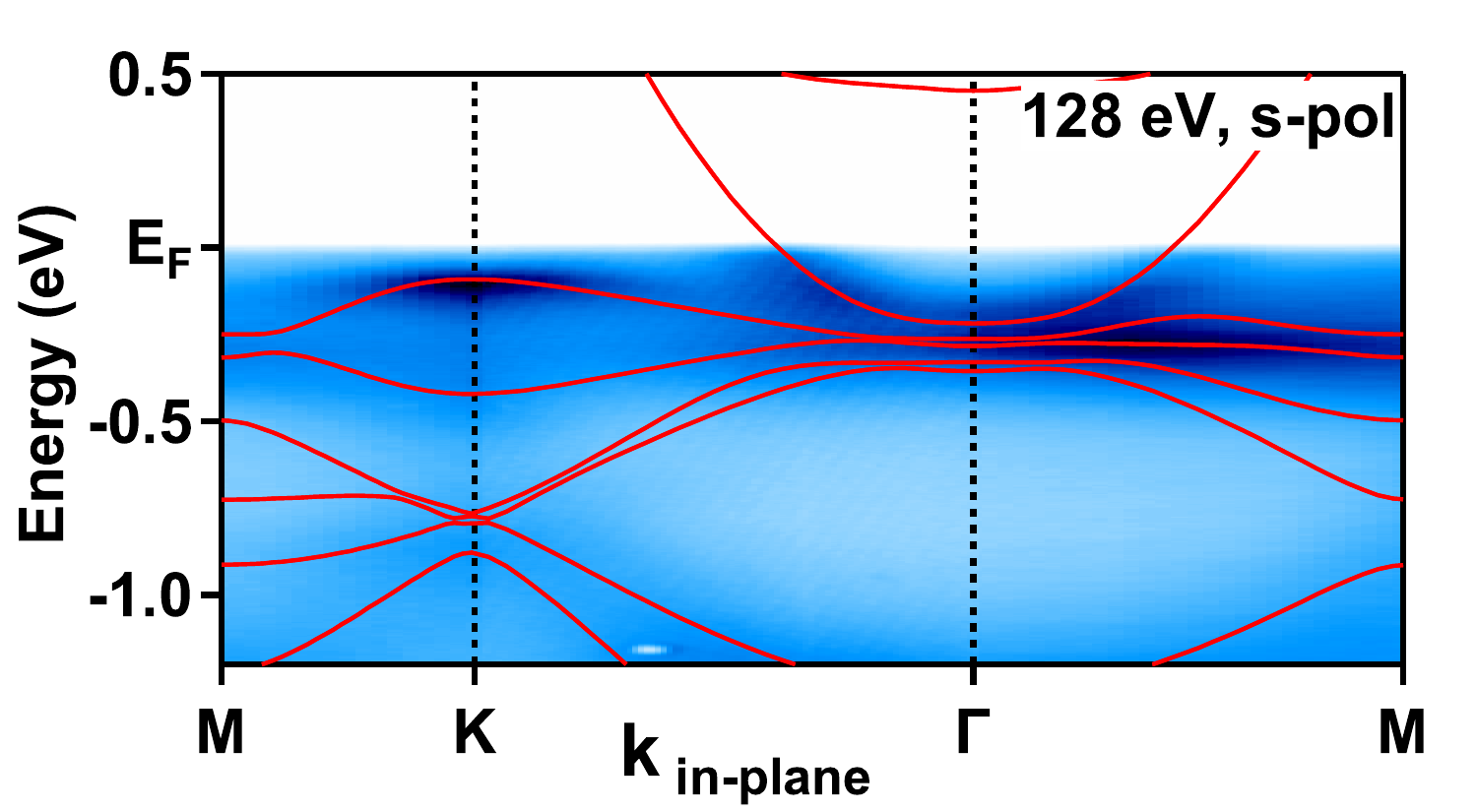}
    }\hfill
    \caption{\label{fig:DFT} Density functional theory calculation results.
    (a) The band structure of pristine CoSn calculated by DFT. The horizontal dashed line represents the Fermi level. FB1 and FB2 are highlighted in blue and green, respectively. 
    (b) The band structure of CoSn with doping of 1.7$\times$10$^{21}$\,electrons/cm$^3$, and binding energy renormalization factor of 0.85. Band I (cyan), II (pink), III (blue), and IV (gold) are the bands crossing the Fermi level.
    (c) Energy shift of FB1 at the H point as a function of doping level.
    (d-e) DFT calculation results (red) overlaid on the ARPES spectra (blue image) taken at (d) $k_z = \pi$ and (e) $k_z = 0$, respectively. The DFT calculation shows good agreement with the experimental data.
    } 
\end{figure}

To investigate the tunability of the flat bands by carrier doping, we performed density functional theory (DFT) calculations (see SI section S1.4 for methods).
Figure~\ref{fig:DFT_prisrtine} shows the band structure of pristine CoSn without carrier doping.
This calculation suggests that FB1 sits exactly at the Fermi level at the H point, while FB2 has a binding energy of -0.26\,eV at the $\Gamma$ point.
We performed additional DFT calculations with different doping concentrations ranging from 1.0$\times$10$^{22}$\,holes/cm$^{3}$ to 1.0$\times$10$^{22}$\,electrons/cm$^{3}$ (see SI section S5).
Within this doping range, adding electrons to the system ($n_{doping}>0$) shifts the FB1 downwards with respect to the Fermi level, while the band dispersion only changes slightly.
The energy shift of FB1 at the H point as a function of doping level is summarized in Figure~\ref{fig:DFT_energy_shift}.
We calculated the band structure with arbitrary doping level by linearly interpolating the DFT results. 
Matching the binding energies of FB1 at the H point and FB2 at the $\Gamma$ point yields a doping level of 1.7$\times$10$^{21}$\,electrons/cm$^3$ and a binding energy renormalization factor of 0.85 (see Figure~\ref{fig:DFT_doped}).
An overlay of DFT bands on top of the ARPES spectra shows a good agreement between them, as shown in Figure~\ref{fig:DFT_stack_97eVLH} and~\ref{fig:DFT_stack_128eVLH}.

\begin{figure}[h]
    \subfloat[\label{fig:Rxx}]{
    \includegraphics[width=0.48\textwidth]{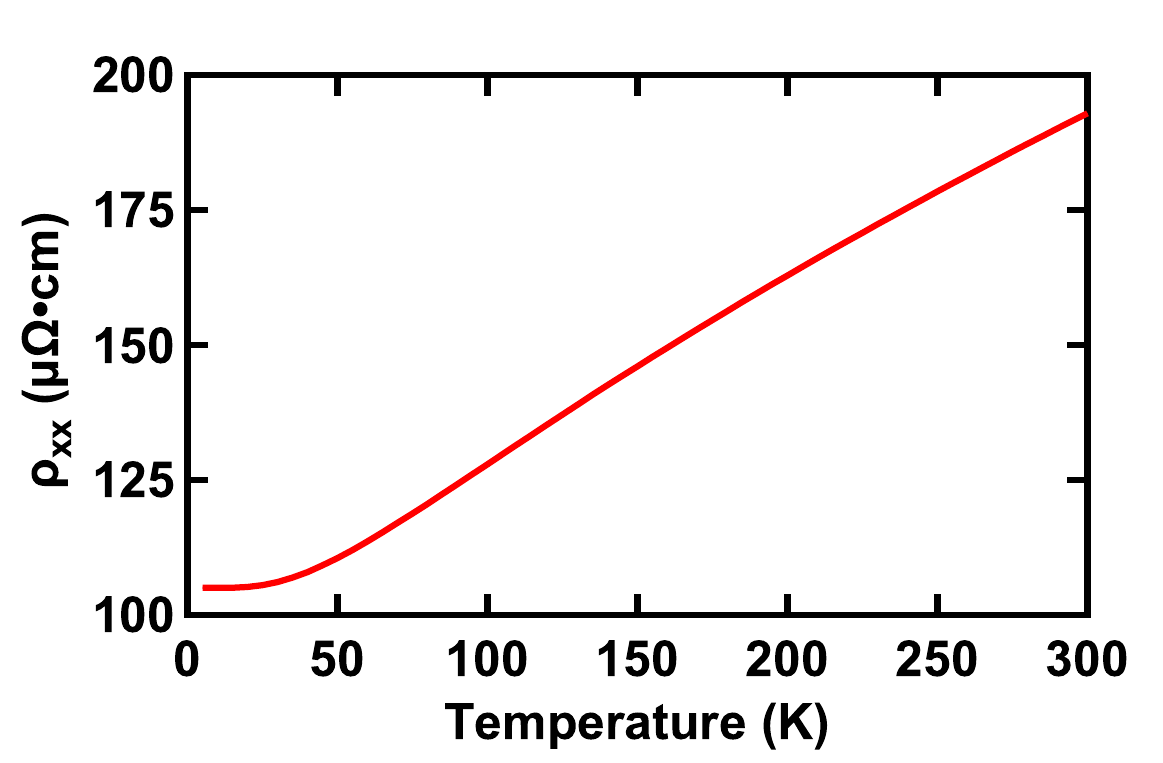}
    }
    \subfloat[\label{fig:Hall}]{
       \includegraphics[width=0.48\textwidth]{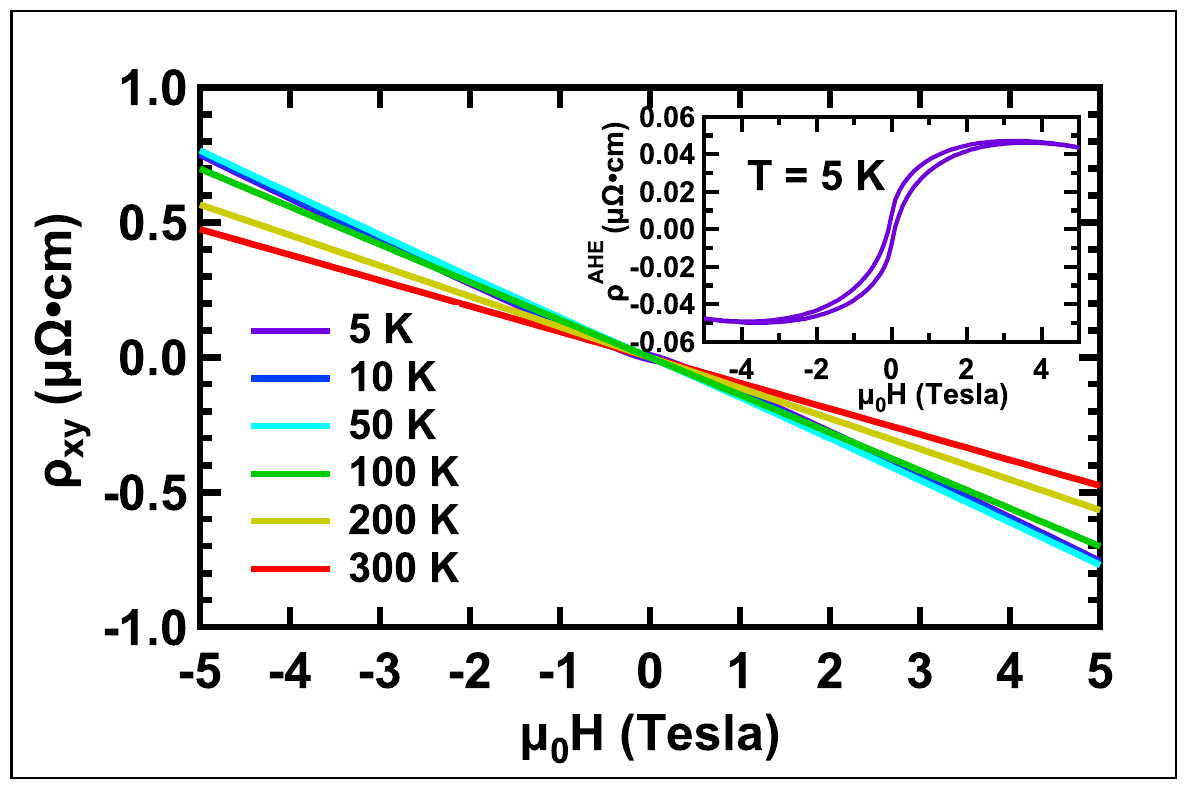}
    }\hfill
    \caption{\label{fig:Transport} Transport properties of a 35\,nm CoSn thin film on MgO(111) substrate.
    (a) Longitudinal resistivity of the CoSn thin film as a function of temperature.
    (b) Hall resistivity of the CoSn thin film at different temperatures.
    } 
\end{figure}

Figure~\ref{fig:Rxx} shows the longitudinal resistivity of a 35\,nm CoSn thin film grown on MgO(111) substrate as a function of temperature (see SI section 1.5 for methods).
At room temperature, the CoSn thin film has a resistivity of 192\,$\mu\Omega\cdot$cm.
As the temperature decreases from room temperature, the resistivity drops almost linearly down to $\sim$30\,K, then reaches a plateau of 105\,$\mu\Omega\cdot$cm.
The Hall resistivity of the same sample at different temperatures is shown in Figure~\ref{fig:Hall}.
At room temperature, the Hall resistivity shows linear behavior with respect to the magnetic field, with no detectable anomalous Hall effect.
This observation is in agreement with the non-magnetic nature of CoSn~\cite{allred2012ordered}.
However, as the temperature drops below 10\,K, an anomalous Hall effect starts to appear, although the amplitude is no larger than 0.05\,$\mu\Omega\cdot$cm (See inset of Figure~\ref{fig:Hall}).
This tiny anomalous Hall effect was also observed in sputtered thin films~\cite{thapaliya2021high} and similar signals were observed in low-temperature magnetization measurements of bulk crystals~\cite{liu2020orbital}.
The mechanism behind it is suggested to be either a small deviation in stoichiometry or magnetism induced by the flat band~\cite{thapaliya2021high}.
The extracted carrier density is 6.57$\times10^{21}$\,electrons/cm$^{3}$ at room temperature and 3.91$\times10^{21}$\,electrons/cm$^{3}$ at 5\,K, respectively.

The relationship between the transport properties and the band structure of CoSn has been discussed qualitatively~\cite{kang2020topological, liu2020orbital, huang2022flat}, however, a quantitative discussion has been missing.
Meanwhile it is important to quantitatively understand the relationship between the transport properties and the band structure of CoSn, especially to disentangle the contribution from individual bands, since the transport measurements provide indirect evidence for flat bands near the Fermi level~\cite{sales2021tuning, huang2022flat}.
Here we provide an estimate of Hall resistivity from DFT-calculated band structures.
In semiclassical transport theory, the contribution to longitudinal conductivity $\sigma_{xx}$ from band $i$ is given by~\cite{hurd1972hall}:
\begin{equation}\label{eq:equation1}
\sigma_{xx}^{(i)}=-\frac{e^2}{4\pi^3\hbar^2}\int\tau_{i}(\textbf{k})\left(\frac{\partial\varepsilon}{\partial k_{x}}\right)^{2}\frac{\partial f}{\partial \varepsilon}d\textbf{k}
\end{equation}
where $\tau_{i}$ is the carrier relaxation time, $\varepsilon(\textbf{k})$ is the energy dispersion, and $f$ is the distribution function.
Meanwhile, the contribution to Hall conductivity $\sigma_{H}$ from band $i$ is given by~\cite{hurd1972hall}: 
\begin{equation}\label{eq:equation2}
\sigma_{H}^{(i)}=-\frac{e^3}{4\pi^3\hbar^4}\int\tau_{i}(\textbf{k})\left(\frac{\partial\varepsilon}{\partial k_{x}}\right)\nabla_{\textbf{k}}\varepsilon\times\nabla_{\textbf{k}}\tau(\textbf{k})\left(\frac{\partial\varepsilon}{\partial k_{y}}\right)\frac{\partial f}{\partial \varepsilon}d\textbf{k}
\end{equation}
Since only partially occupied bands contribute to the transport at low temperatures, we hereby calculated the $\sigma_{xx}^{(i)}/\tau_{i}$ and $\sigma_{H}^{(i)}/\tau_{i}^2$ for the bands crossing the Fermi level, namely band I (cyan), II (pink), III (blue), and IV (gold) in Figure~\ref{fig:DFT_doped} (see SI section S6 for details).
Assuming a universal and k-independent relaxation time $\tau$ for all bands, we calculated the Hall coefficient $R_{H}$ by~\cite{hurd1972hall}:
\begin{equation}\label{eq:equation3}
R_H=\frac{\sigma_{H}}{\sigma_{xx}^2}=\frac{\sum_{i}\sigma_{H}^{(i)}}{(\sum_{i}\sigma_{xx}^{(i)})^{2}}
\end{equation}
which is independent of relaxation time $\tau$.
With this method, our theory gives a Hall resistivity of -1.0$\times$10$^{-9}$\,m$^3$/C, which has the same order of magnitude as the experimental value of -1.6$\times$10$^{-9}$\,m$^3$/C at T = 5\,K.
The major contribution to the Hall coefficient comes from band I (see Table~S2 in the SI), which has a large electron pocket around the $\Gamma$ point.
This analysis also yields values for $\sigma_{xx}$ that are consistent with our experimental results if we assume reasonable values for $\tau$, as discussed in SI section S6.
Our theoretical modeling bridges the gap between the band structure and the experimentally measured transport properties of CoSn.

In conclusion, we synthesized the epitaxial CoSn thin films directly on insulating substrates and studied their electronic band structures. 
The three-step growth generated highly ordered CoSn (0001) thin films, as confirmed by a combination of RHEED, XRD, and STEM.
The electronic band structures of CoSn thin films were measured with synchrotron-based ARPES.
The flat bands were clearly visualized and the topologically non-trivial nature of the flat band is signified by the spin-orbit coupling gap at the band touching point.
The ARPES measurement, DFT calculations, and the transport properties of CoSn are consistent with each other not only qualitatively but also quantitatively.
One very interesting direction in the near future will be the fabrication of devices that allows voltage gating in order to tune the flat bands across the Fermi level.
This work makes the epitaxial CoSn thin films ready for studies of strongly correlated electronic states and flat band-induced phenomena.

\section*{Methods}
  
  See SI section S1.

\begin{suppinfo}

Methods (MBE, STEM, ARPES, DFT, transport measurements), RHEED and XRD of CoSn on MgO(111), additional STEM data, additional ARPES data, additional DFT calculations, and estimation of longitudinal conductivity and Hall conductivity.

\end{suppinfo}

\begin{acknowledgement}

We thank Dr.~Tiancong Zhu for technical assistance.
This work was supported by NSF Grant No.~1935885, AFOSR MURI 2D MAGIC Grant 
No.~FA9550-19-1-0390, U.S. DOE Office of Science, Basic Energy Sciences Grants No.~DE-SC0016379 and No.~DE-SC0004890, the Center for Emergent Materials, an NSF MRSEC, under Grant No.~DMR-2011876, and the UB Center for Computational Research. This research used resources of the Advanced Light Source, which is a DOE Office of Science User Facility under contract No.~DE-AC02-05CH11231. Electron microscopy was performed at the Center for Electron Microscopy and Analysis (CEMAS) at The Ohio State University.

\end{acknowledgement}

\section*{Author Contributions Statement}

S.C., M.N., W.Z., A.J.B., C.J., A.B., and E.R. performed ARPES measurements.
S.C. and R.K.K. analyzed the ARPES data.
S.C. performed MBE growth, transport measurements, and theoretical modeling of transport data.
T.Z. and I.\v{Z} performed DFT calculations.
N.B. and D.W.M. performed STEM measurements.
I.L. performed XRD measurements and fabricated devices.
S.C. and R.K.K. conceived the project and R.K.K. supervised the project. All authors contributed to writing the manuscript.

\bibliography{CoSn.bib}

\newpage

\section*{TOC Graphic}

\begin{figure}[h]
    {\label{fig:ToC}
    \includegraphics[width=\textwidth]{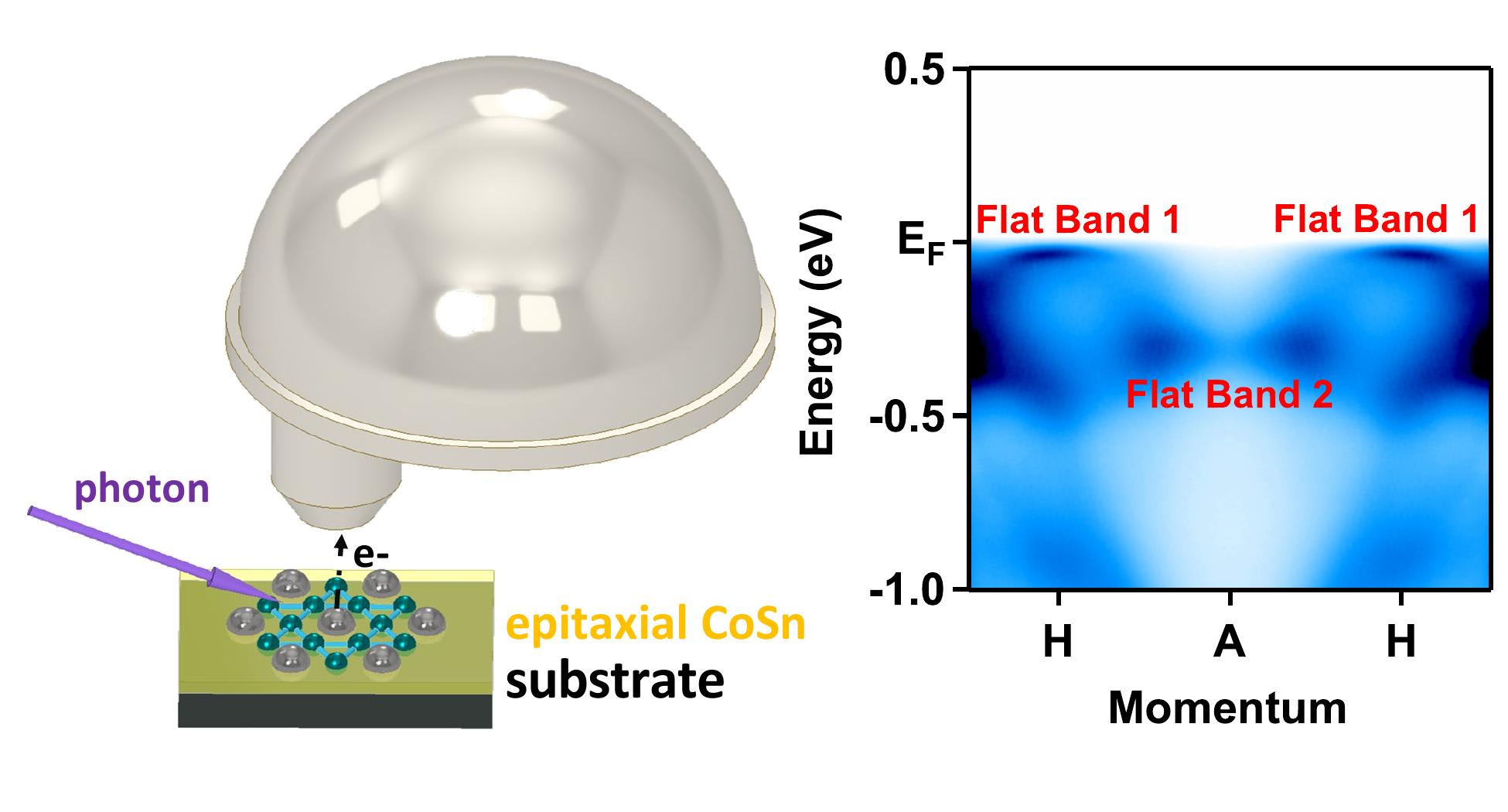}
    } 
\end{figure}

\end{document}

\maketitle

\newpage

\tableofcontents

\newpage

\section{Methods}

 \subsection{Molecular Beam Epitaxy Growth of CoSn Thin Films}

The epitaxial growth of CoSn thin films was performed in an ultra-high vacuum chamber with a base pressure of 2$\times10^{-9}$\,Torr.
Prior to the growth, the MgO(111) or 4H-SiC(0001) substrates were degassed \textit{in-situ} at 500\,$^\circ$C for 20 minutes.
After annealing, the nucleation layer of 5\,nm CoSn was grown at 500\,$^{\circ}$C on 4H-SiC(0001) substrates, or at 470\,$^{\circ}$C on MgO(111) substrates.
After the growth of the nucleation layer, the sample was allowed to cool down to 100\,$^{\circ}$C in 60 minutes.
A $15-20$\,nm continuation layer was grown at 100\,$^{\circ}$C, followed by a temperature ramp to 300\,$^{\circ}$C at 12\,$^{\circ}$C/min.
Finally, a $5-10$\,nm terminating layer was grown on top of the continuation layer at 300\,$^{\circ}$C to improve the surface crystallinity.
The Co and Sn materials were evaporated from Knudson cells, and the typical growth rates were 0.78\,\AA/min~for Co, and 1.91\,\AA/min~for Sn, respectively.
The deposition rates were measured using a quartz crystal deposition monitor.

  \subsection{Scanning Transmission Electron Microscopy (STEM)}

The cross-sectional STEM samples were prepared by Ga ion milling using an FEI Helios Nano Lab 600 Dual Beam focused ion beam (FIB) operated at 30\,kV and 5\,kV.  
Final cleaning passes to remove any amorphous damage layers created in the FIB were performed in a Fischione Nanomill with 900\,V and then 500\,V Ar ions at Cryogenic temperature. 

HAADF-STEM imaging was performed using a probe-corrected Themis-ZTM at 300\,kV. 
Images were collected using the drift-corrected-frame-integration (DCFI) acquisition method within the Thermo Scientific Velox Software.

   \subsection{Angle-Resolved Photoemission Spectroscopy Measurements}

The ARPES experiments were performed at Beamline 7.0.2 (MAESTRO) of the Advanced Light Source (ALS). 
The samples were transferred from the growth chamber into the ultra-high vacuum (UHV) suitcase and shipped to the ALS, then transferred from the UHV suitcase into the ARPES system.
During the entire sample transfer procedure, the samples were kept under vacuum without exposure to air, to ensure a clean surface for ARPES experiments.
The ARPES experiments were performed at T = 6\,K. 
The photoemitted electrons were collected using a Scienta Omicron R4000 hemispherical electron analyzer, which provides energy and momentum resolution better than 30\,meV, and 0.01\,\AA$^{-1}$, respectively.

  \subsection{Density Functional Theory (DFT) Calculations}

The geometry optimization and electronic structure calculations were performed using the first-principles method based on density functional theory (DFT) with the projector-augmented-wave (PAW) formalism, as implemented in the Vienna \textit{ab initio} simulation package (VASP)~\cite{kresse1996efficient}. 
All calculations were carried out with a plane-wave cutoff energy of 550\,eV and 15$\times$15$\times$11 Monkhorst-Pack grids were adopted for the first Brillouin zone integral. 
The Perdew-Burke-Ernzerhof generalized-gradient approximation (GGA) was used to describe the exchange and correlation functional~\cite{perdew1996generalized}. 
The convergence criterion for the total energy is 10$^{-6}$\,eV. 
All the atoms in the unit cell are allowed to move until the Hellmann-Feynman force on each atom is smaller than 0.01\,eV/\AA. 
The lattice constants of CoSn are a = b = 5.275\,\AA~and c = 4.263\,\AA, taken from the experiments.

  \subsection{Transport Measurements}

The transport measurements were performed using a Quantum Design physical properties measurement system (PPMS) in DC resistivity mode.
The 35\,nm CoSn sample was patterned into 100\,$\mu$m-wide, 300\,$\mu$m-long Hall bar devices.
The longitudinal resistance was measured using 4-probe geometry.

\newpage

\section{RHEED and XRD of CoSn on MgO(111)}

\begin{figure}
    \subfloat[\label{fig:RHEED_MgO}]{
    \includegraphics[width=0.42\textwidth]{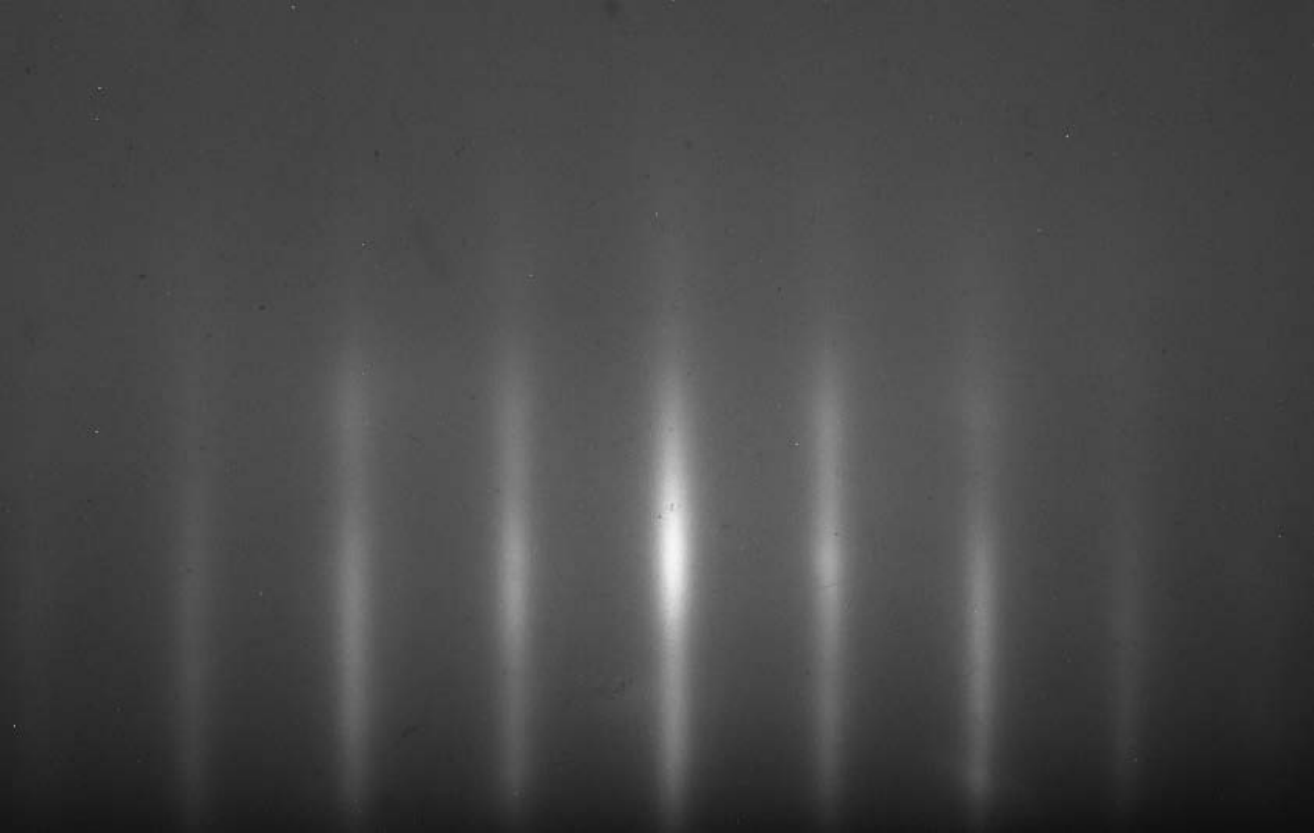}
    }
    \subfloat[\label{fig:XRD_MgO}]{
       \includegraphics[width=0.58\textwidth]{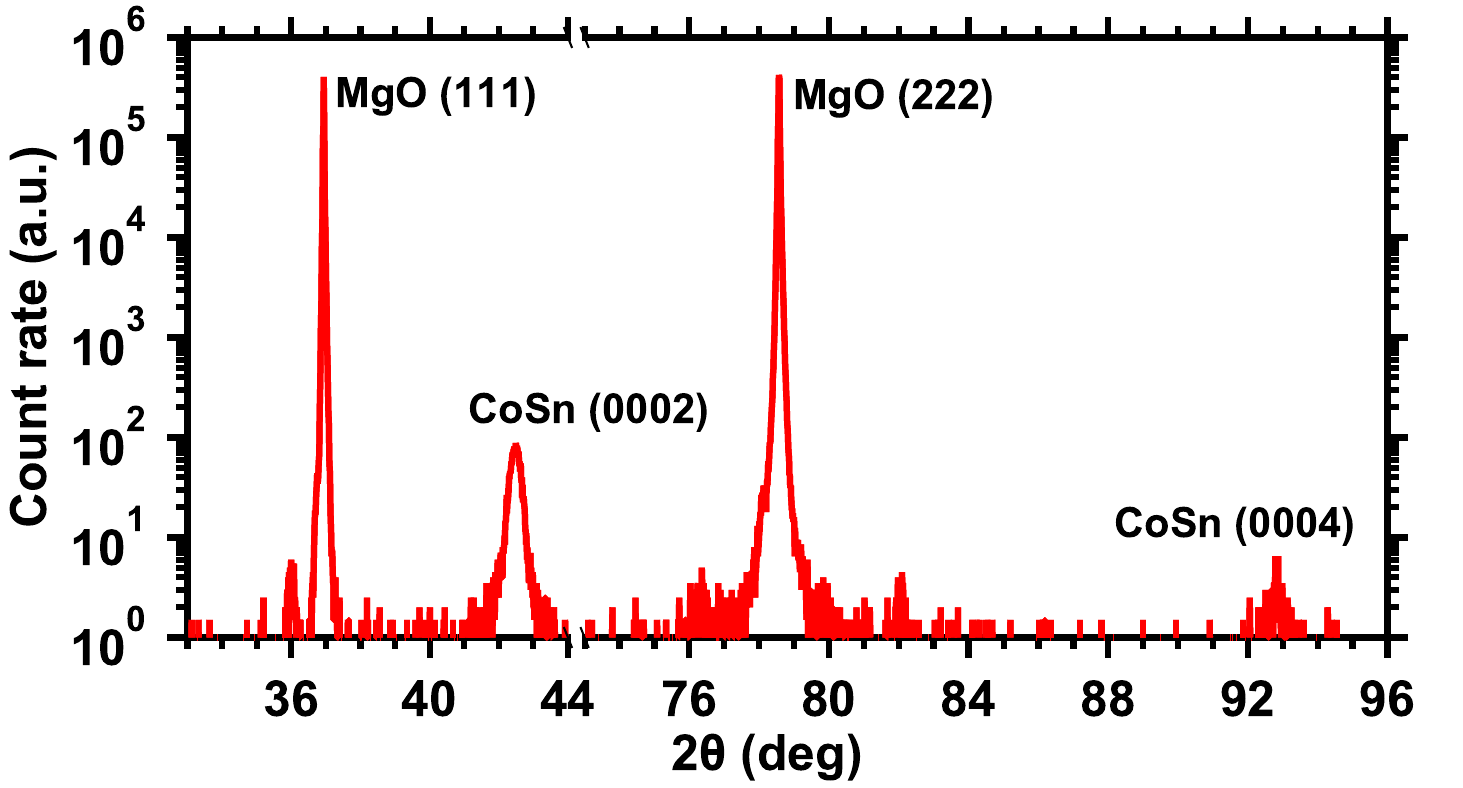}
    }\hfill
    \caption{\label{fig:CoSn_onMgO} Growth of 35\,nm CoSn thin films grown on MgO(111) substrates.
    (a) \textit{In-situ} RHEED pattern of a 35\,nm CoSn thin film grown on MgO(111) substrate.
    (b) XRD data of a 35\,nm CoSn thin film grown on MgO(111) substrate.
    } 
\end{figure}

Figure~\ref{fig:RHEED_MgO} shows the \textit{in-situ} RHEED pattern of a 35\,nm CoSn sample grown on MgO(111) substrate.
The streaky RHEED pattern indicates epitaxial growth and two-dimensional surfaces with finite terrace width.
Figure~\ref{fig:XRD_MgO} shows the XRD data of the 35nm\,CoSn sample grown on MgO(111) substrate.
CoSn (0002) and (0004) peaks show up at 42.49\,$^\circ$ and 92.87\,$^\circ$, respectively.
The out-of-plane lattice constant extracted from the XRD scan is 4.252\,\AA, which is nearly identical to the lattice constant of CoSn grown on 4H-SiC(0001) substrates, and consistent with the previously reported values from bulk crystals~\cite{sales2021tuning} and sputtered thin films~\cite{thapaliya2021high}.

\newpage

\section{Additional STEM Data}

\begin{figure}[h]
    \includegraphics[width=\textwidth]{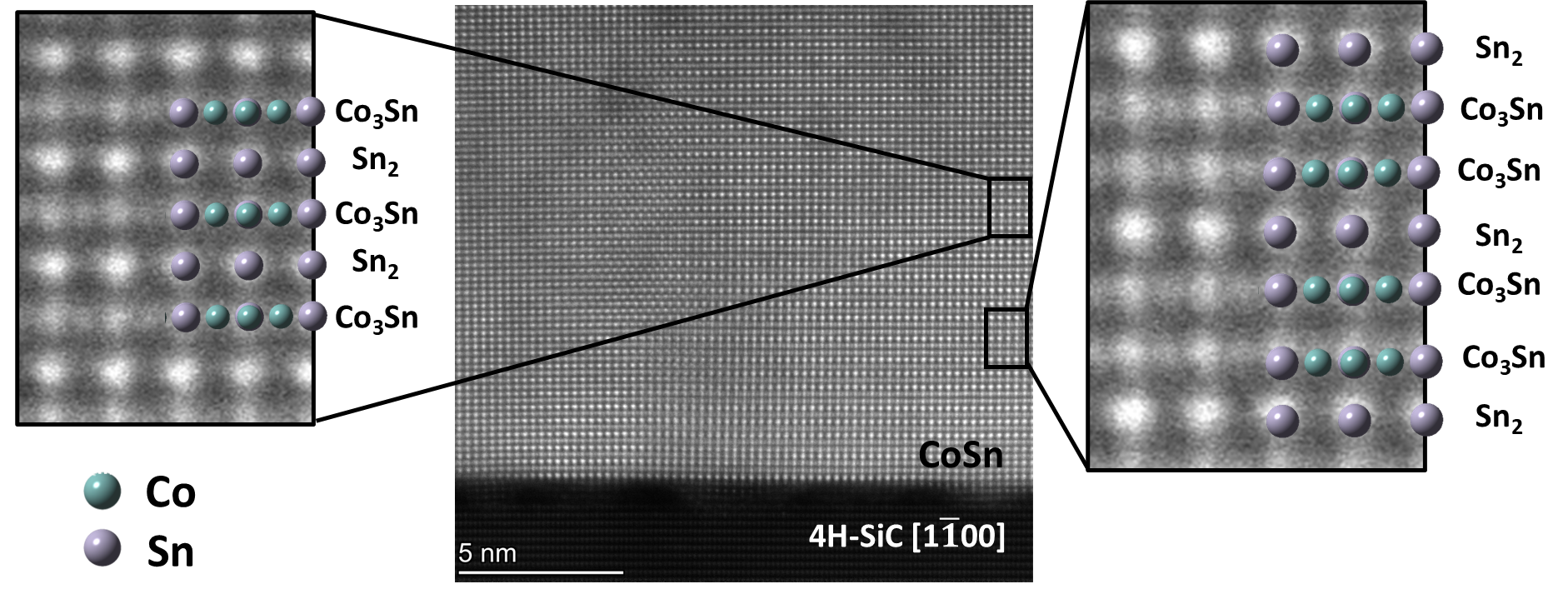}
    \caption{\label{fig:STEM_Supp} 
    Atomic-resolution HAADF-STEM image of a CoSn(0001) thin film grown on 4H-SiC(0001) substrate viewed along 4H-SiC[1$\bar{1}$00] direction. Left panel: zoomed-in image of CoSn with expected stacking sequence of one Co$_3$Sn layer followed by one Sn$_2$ layer. Right panel: zoomed-in image of CoSn with additional Co$_3$Sn layers near the interface.
    } 
\end{figure}

Figure~\ref{fig:STEM_Supp} shows the atomic-resolution HAADF-STEM image of a CoSn(0001) thin film grown on 4H-SiC(0001) substrate viewed along 4H-SiC[1$\bar{1}$00] direction.
Across the film, the predominant stacking sequence is alternating stacking of one Co$_3$Sn layer and one Sn$_2$ layer, which is expected for CoSn (see the left panel of Figure~\ref{fig:STEM_Supp}).
However, in some regions near the interface, the stacking fault happens, resulting in a stacking sequence with additional Co$_3$Sn layers, as shown in the right panel of Figure~\ref{fig:STEM_Supp}.
Additionally, several-nanometer surface steps, the vertical shift of layer sequence, distortions, and contrast variations of the lattice are observed across the film suggesting the presence of defects and a complex nanostructure.

Using the lattice constant of the substrate as a reference for calibration, fast Fourier transform (FFT) analysis of the STEM image over a few nanometer scale yields lattice constants of $a$ = 5.36$\pm$0.24\,\AA~and $c$ = 4.25$\pm$0.12\,\AA~for CoSn.
The out-of-plane lattice constant $c$ agrees well with our XRD result of $c$ = 4.254\,\AA, and is consistent with previously reported values from bulk crystals~\cite{sales2021tuning} and thin films~\cite{thapaliya2021high}.
Meanwhile, the in-plane lattice constant $a$ is 1.5\% larger than the bulk value of $a$ = 5.279\,\AA~\cite{sales2021tuning}, but the difference is within the uncertainty of FFT results. 
Previous DFT calculations suggest that tensile strain shifts the flat bands downwards with respect to the Fermi level~\cite{kang2020topological}.
Along the direction of growth, there is no obvious change in the lattice constants observed in STEM, suggesting that the strain is relaxed within the first few nanometers of growth.

\newpage

\section{Additional ARPES Data}

    \subsection{ARPES Spectrum of the CoSn/MgO(111) Sample}

\begin{figure}[htbp]
    \subfloat[\label{fig:FB_97eVLH_MgO}]{
    \includegraphics[width=0.48\textwidth]{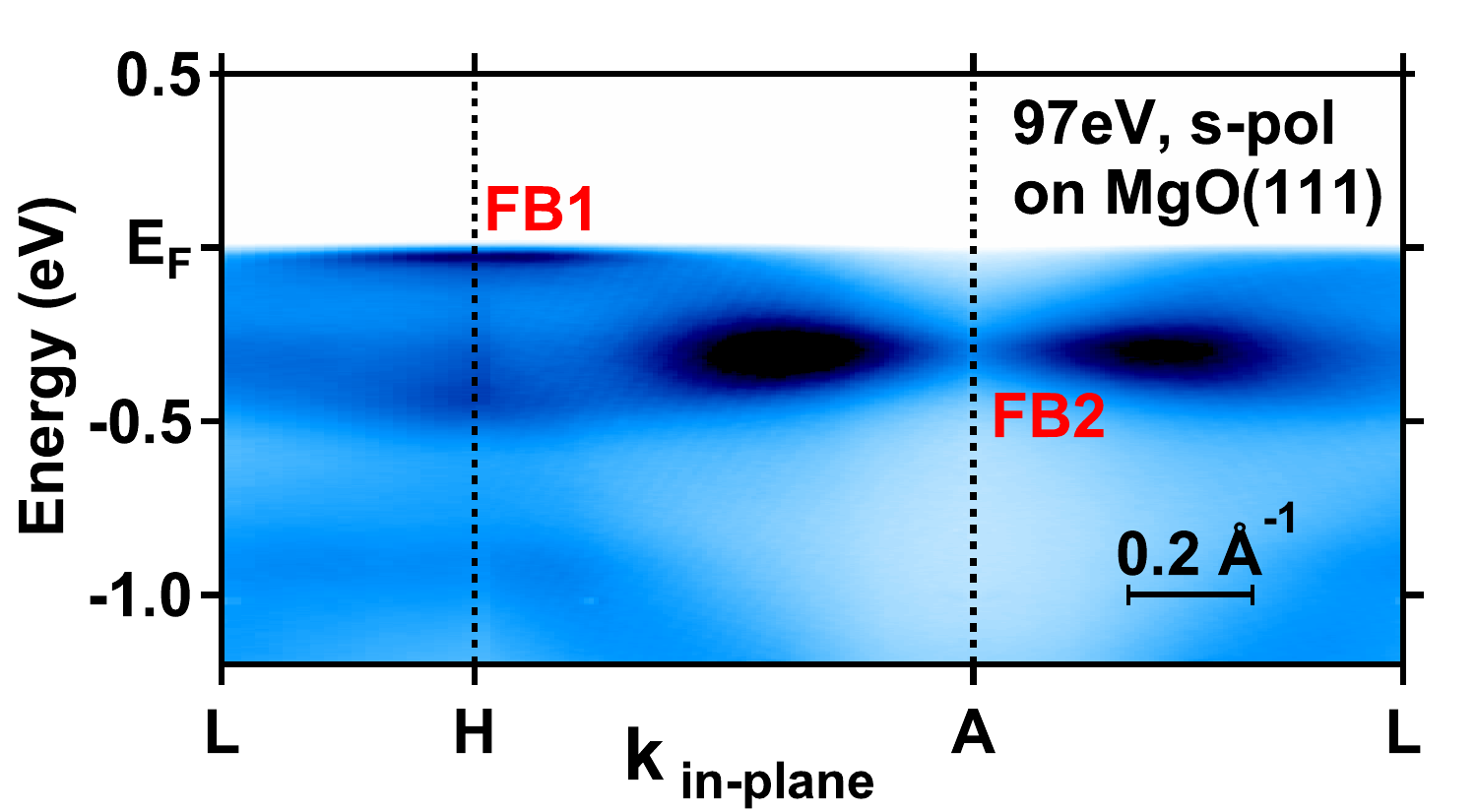}
    }
    \subfloat[\label{fig:FB_128eVLH_MgO}]{
       \includegraphics[width=0.48\textwidth]{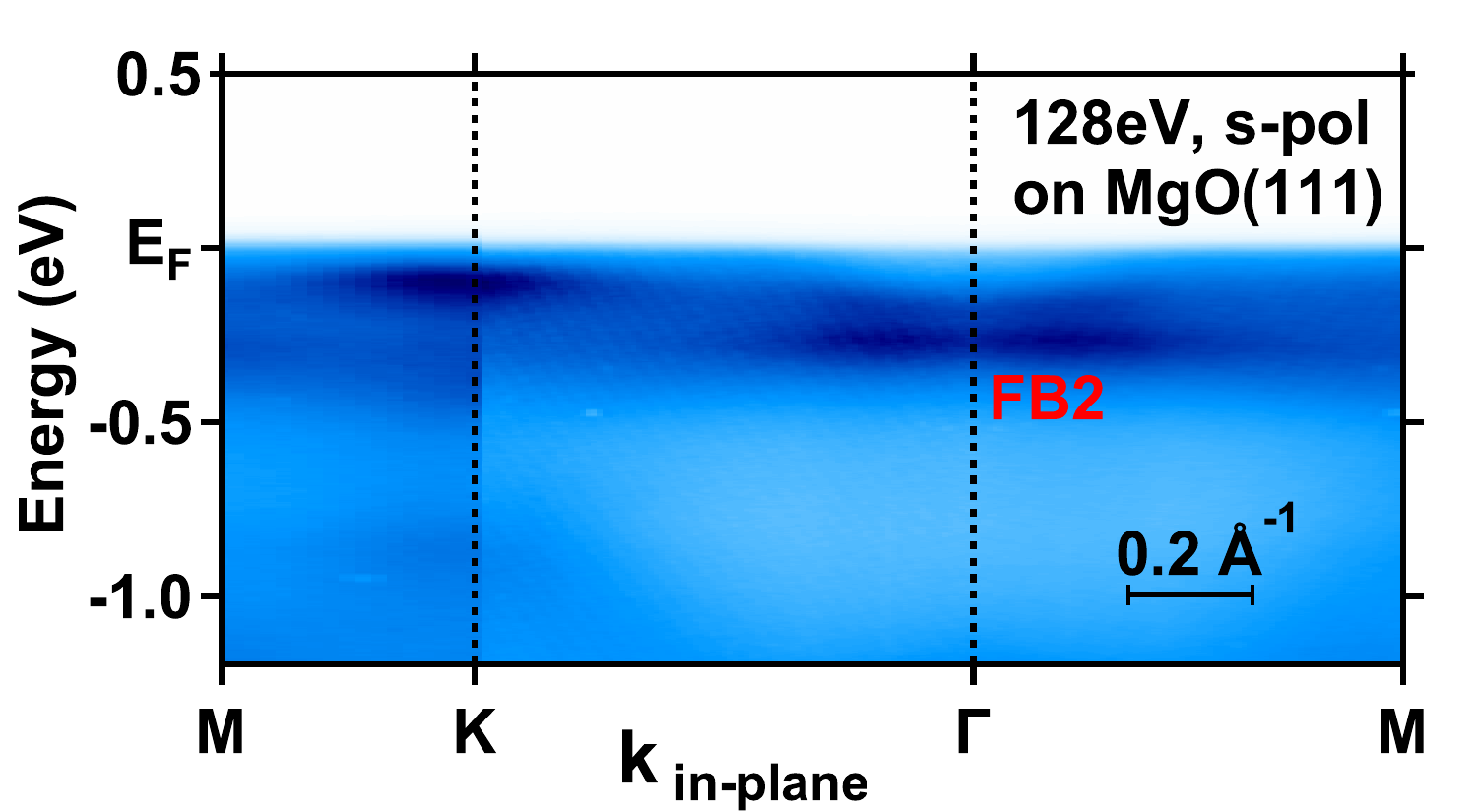}
    }\hfill
    \caption{\label{fig:ARPES_MgO} ARPES spectrum of the CoSn/MgO(111) sample taken using (a) 97\,eV s-polarized photons (corresponding to $k_z=\pi$ (mod 2$\pi$) plane) and (b) 128\,eV s-polarized photons (corresponding to $k_z=0$ (mod 2$\pi$) plane), respectively.
} 
\end{figure}

Figure~\ref{fig:ARPES_MgO} shows the ARPES spectrum of the CoSn(0001) thin film grown on MgO(111) substrate.
Generally, the spectrum exhibits similar features as compared to the CoSn(0001) thin film grown on 4H-SiC(0001) substrates.
In the $k_z=\pi$ (mod 2$\pi$) plane, FB1 is observed around the H points, 0.03\,eV below the Fermi level, while FB2 spreads over almost the entire BZ with a binding energy of about -0.3\,eV (Figure~\ref{fig:FB_97eVLH_MgO}).
In the $k_z=0$ (mod 2$\pi$) plane, only FB2 is observed at 0.3\,eV below Fermi level.
Compared to the CoSn/4H-SiC(0001) sample, the differences in binding energies of individual flat bands are within the resolution of the experiment ($\sim$15\,meV).

    \subsection{k$_z$ Dependence of ARPES Spectrum}

\begin{figure}[htbp]
    \includegraphics[width=0.5\textwidth]{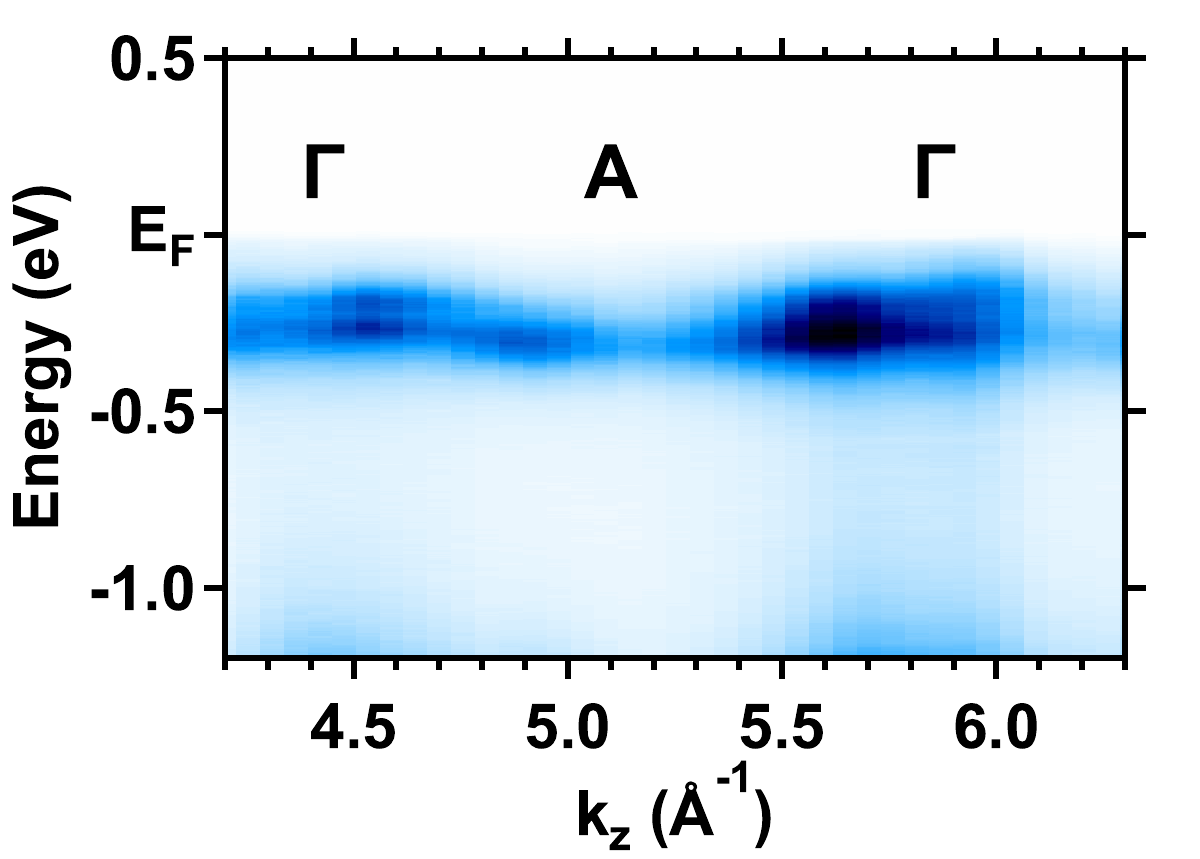}
    \caption{\label{fig:kz} k$_z$ dependence of ARPES spectrum along A-$\Gamma$-A direction using s-polarized photons.
} 
\end{figure}

To determine the high-symmetry planes of the BZ, we measured the ARPES spectrum along the $\Gamma$-A-$\Gamma$ direction by varying the photon energies, as shown in Figure~\ref{fig:kz}.
Along $\Gamma$-A-$\Gamma$ direction, the binding energy of FB2 shows periodic variation, with highest energies corresponding to $\Gamma$ points, and lowest energies corresponding to A points~\cite{kang2020topological}.

    \subsection{Polarization Dependence of ARPES Spectrum}

\begin{figure}[h]
    \subfloat[\label{fig:FB_97eVLH_S}]{
    \includegraphics[width=0.51\textwidth]{Figure_FB/FlatBand_97eVLH.pdf}
    }
    \subfloat[\label{fig:FB_128eVLH_S}]{
       \includegraphics[width=0.46\textwidth]{Figure_FB/FlatBand_128eVLH.pdf}
    }\hfill
    \subfloat[\label{fig:FB_97eVLV}]{
    \includegraphics[width=0.46\textwidth]{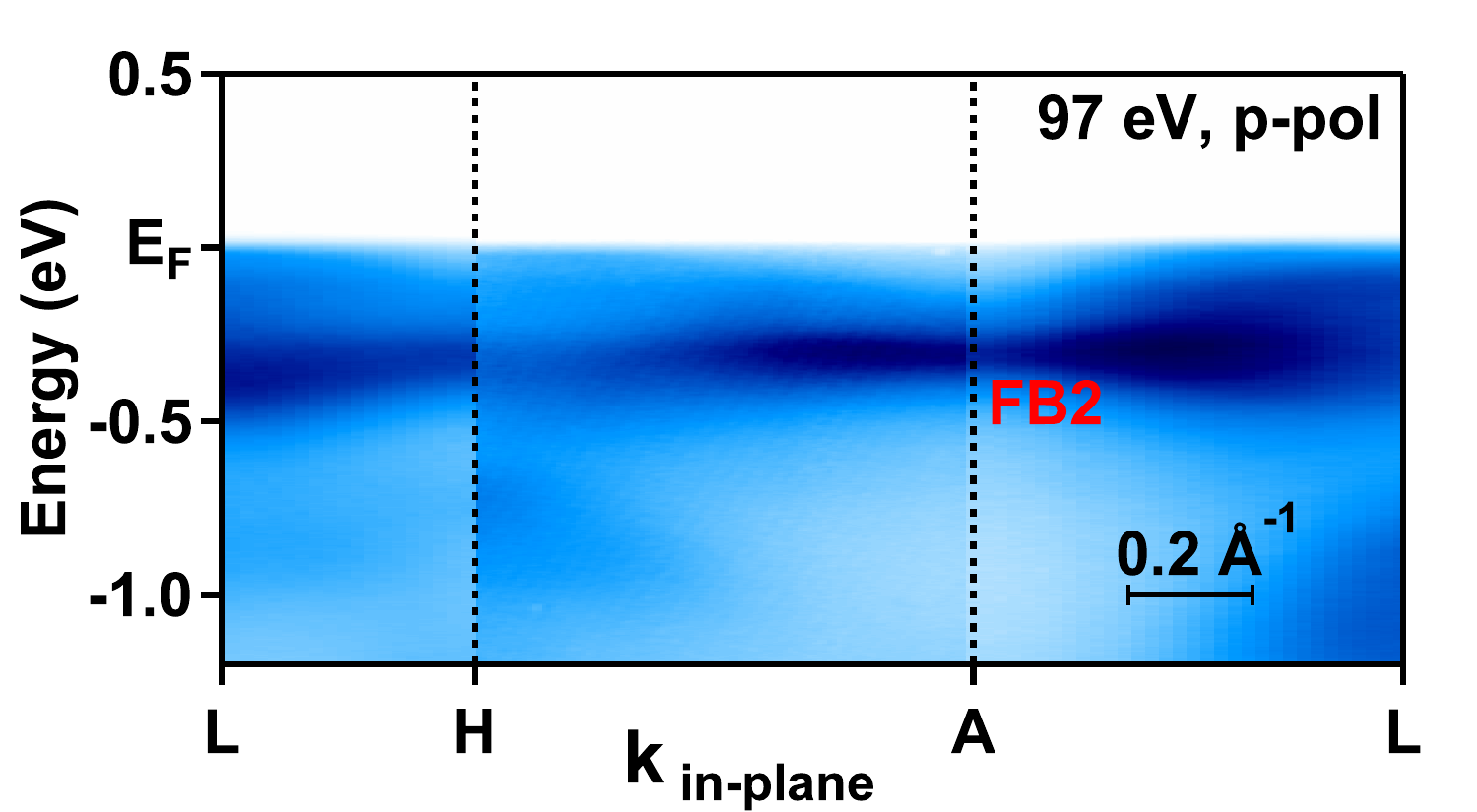}
    }\hfill
    \subfloat[\label{fig:FB_128eVLV}]{
       \includegraphics[width=0.46\textwidth]{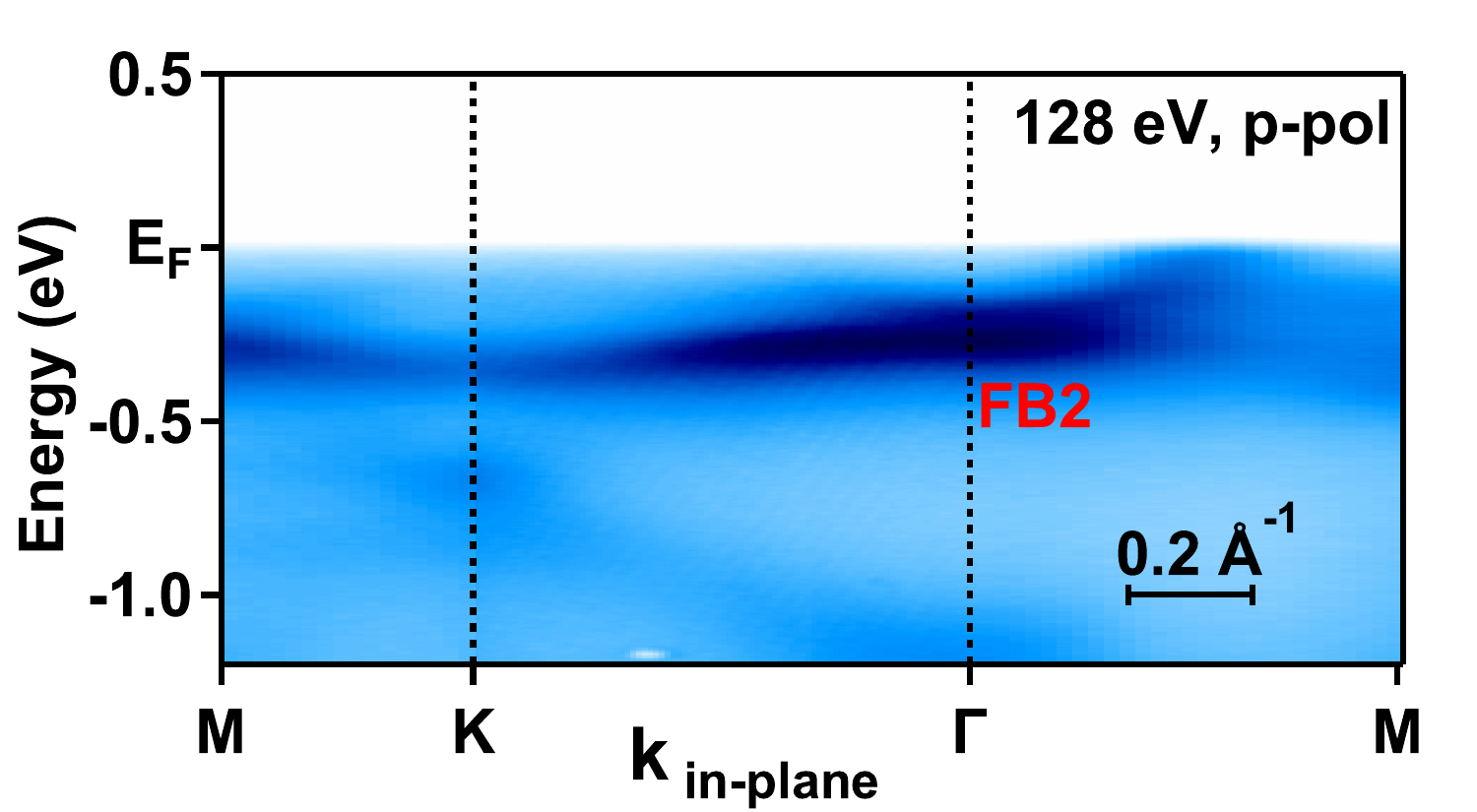}
    }\hfill
    \caption{\label{fig:Polarization} Polarization dependence of ARPES spectrum.
    (a - b) ARPES spectrum taken using s-polarized light at (a) $k_z = \pi$ (mod $2\pi$) plane, (b) $k_z = 0$ (mod $2\pi$) plane, respectively. 
    (c - d) ARPES spectrum taken using p-polarized light at (c) $k_z = \pi$ (mod $2\pi$) plane, (d) $k_z = 0$ (mod $2\pi$) plane, respectively. 
} 
\end{figure}
    
Figure~\ref{fig:Polarization} shows the polarization dependence of the ARPES spectrum of CoSn(0001) on SiC(0001).
In the $k_z = \pi$ (mod $2\pi$) plane, FB1 can be seen using s-polarized photons (Figure~\ref{fig:FB_97eVLH_S}), while it is not visible in the spectrum using p-polarized light (Figure~\ref{fig:FB_97eVLV}).
In the $k_z = 0$ (mod $2\pi$) plane, s-polarized light mostly highlights the upper branch of FB2 at the K points (Figure~\ref{fig:FB_128eVLH_S}), while p-polarized light mostly highlights the lower branch of FB2 (Figure~\ref{fig:FB_128eVLV}).
At the A and $\Gamma$ points, FB2 shows negligible polarization dependence. 
  
The polarization dependence of ARPES spectrum originates from the matrix element effect, which is sensitive to the orbital nature of the bands~\cite{hufner2013photoelectron}. 
The observation above is consistent with the DFT results that FB1 and the upper branch of FB2 mainly originate from the $d_{xy}$ and $d_{x^2-y^2}$ orbitals, while the lower branch of FB2 mainly originates from the $d_{xz}$ and $d_{yz}$ orbitals.

    \subsection{``Flatness" of the Flat Bands}

\begin{figure}[h]
    \subfloat[\label{fig:97eV_HLH}]{
    \includegraphics[height = 1.83in]{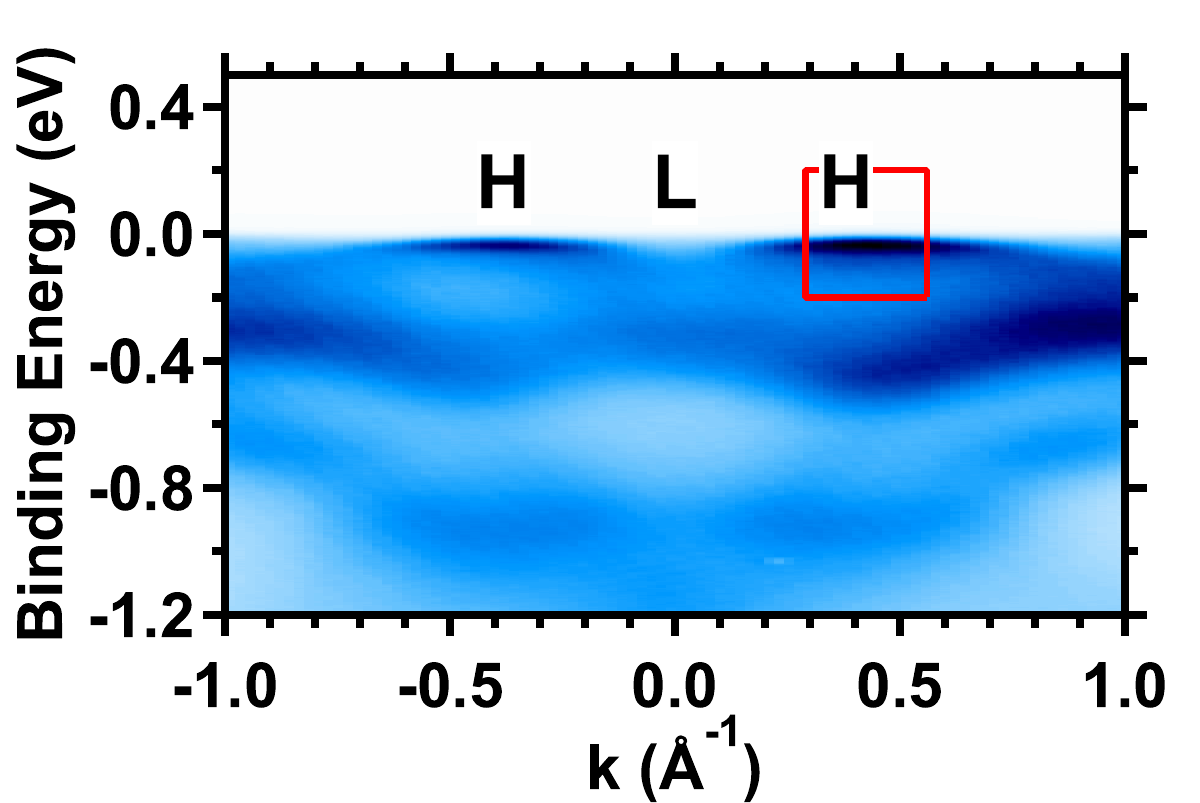}
    }
    \subfloat[\label{fig:EDC_Stack_H}]{
       \includegraphics[height = 1.83in]{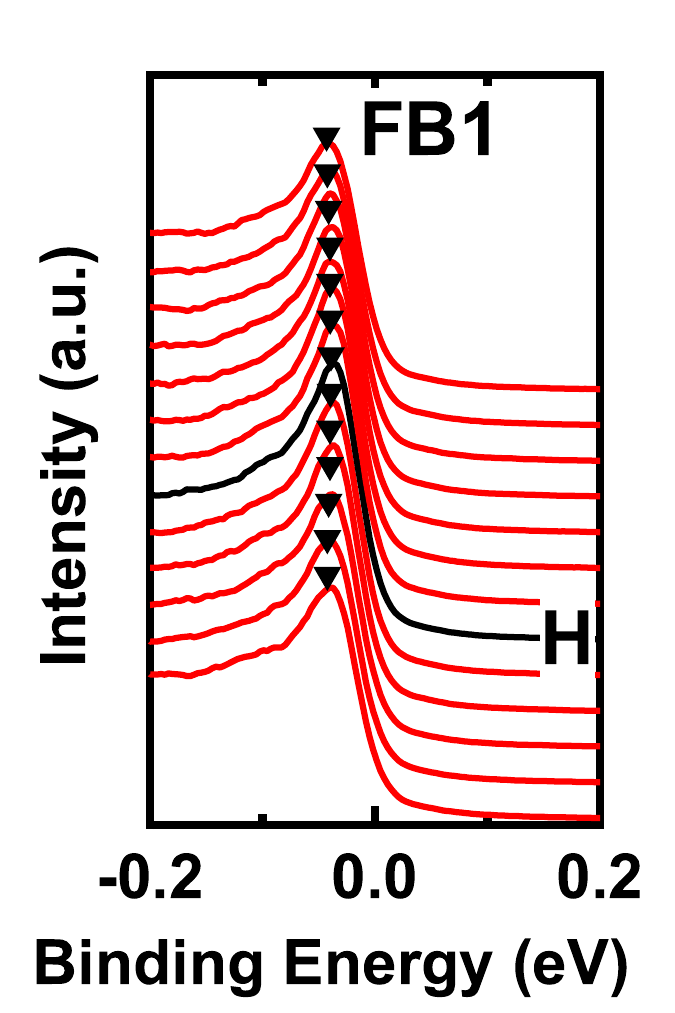}
    }
    \subfloat[\label{fig:EDC_FB2}]{
       \includegraphics[height = 1.83in]{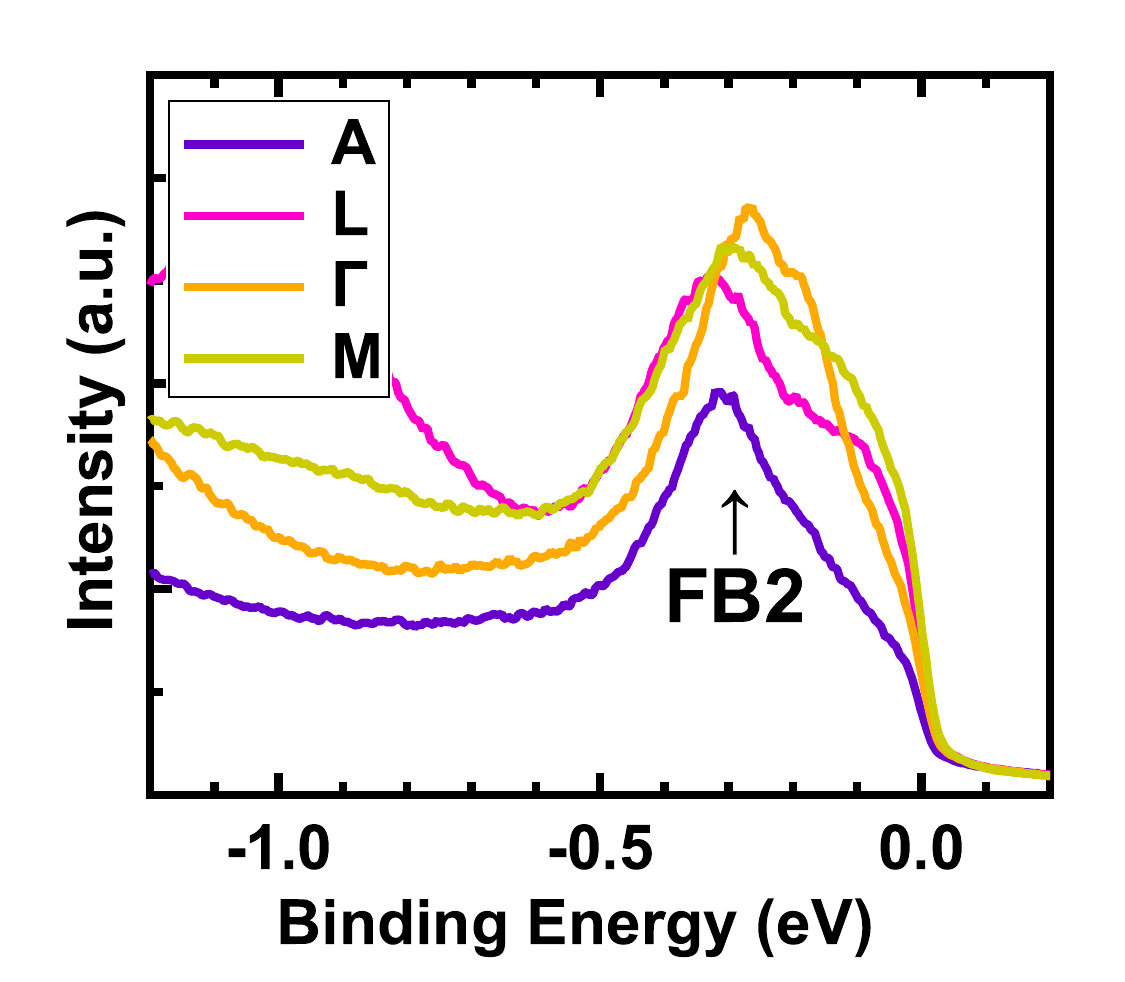}
    }\hfill
    \caption{\label{fig:EDC} 
    (a) Spectrum of CoSn(0001)/SiC(0001) along H-L-H direction.
    (b) Extracted EDCs from the red box in (a). The delta symbols represent the peak positions of FB1 from Lorentzian fits.
    (c) EDCs at A (purple), L (pink), $\Gamma$ (orange), and M (yellow) points, highlighting the peaks at about -0.3\,eV binding energy.
    } 
\end{figure}

To examine the ``flatness" of FB1, we extracted the energy distribution curves (EDCs) from the red box of Figure~\ref{fig:97eV_HLH}.
By fitting the EDCs with the product of the Lorentzian function and Fermi-Dirac distribution, we found that the binding energy of FB1 is around -0.04\,eV, as shown in Figure~\ref{fig:EDC_Stack_H}.
The effective mass $m^{*}$ is then calculated from the formula:
\begin{equation}
\frac{1}{m^{*}}=\frac{1}{\hbar^2}\frac{d^2E}{dk^2} \label{eq1}
\end{equation}
After fitting in the peak positions versus the corresponding momentum with a quadratic relationship, the $m^{*}$ is determined to be 16.7\,$m_0$ along L-H direction, where $m_0$ is the mass of free electrons.
The enhancement of effective mass comes from the dispersionless nature of FB1.

\begin{table*}[h]
\begin{tabular*}{\textwidth}{c @{\extracolsep{\fill}} ccccc}
 Position & A & L & $\Gamma$ & M \\\hline
 Binding Energy (eV) & -0.31 & -0.32 & -0.28 & -0.29 \\
 FWHM (eV) & 0.20 & 0.28 & 0.22 & 0.33 \\
\end{tabular*}
\caption{\label{tab:table1} Summary of binding energies and FWHM of FB2 at high-symmetry points.}
\end{table*}

To examine the ``flatness" of FB2, we took EDCs from representative high-symmetry points, as shown in Figure~\ref{fig:EDC_FB2}.
Using Lorentzian fitting, we extracted the binding energies and the full width at half maximum (FWHM) of FB2 at the  A, L, $\Gamma$, and M point, and summarize them in Table~\ref{tab:table1}.
At all these points, FB2 is located at $\sim$0.3\,eV below the Fermi level, and has an FWHM ranging between 0.20\,eV and 0.33\,eV.
The variation of binding energy is within 0.03\,eV, signifying that the FB2 is non-dispersive across almost the entire BZ.

\newpage

\section{Additional DFT Calculations}

\begin{figure}[H]
    \subfloat[\label{fig:DFT_1hole}]{
    \includegraphics[width=0.49\textwidth]{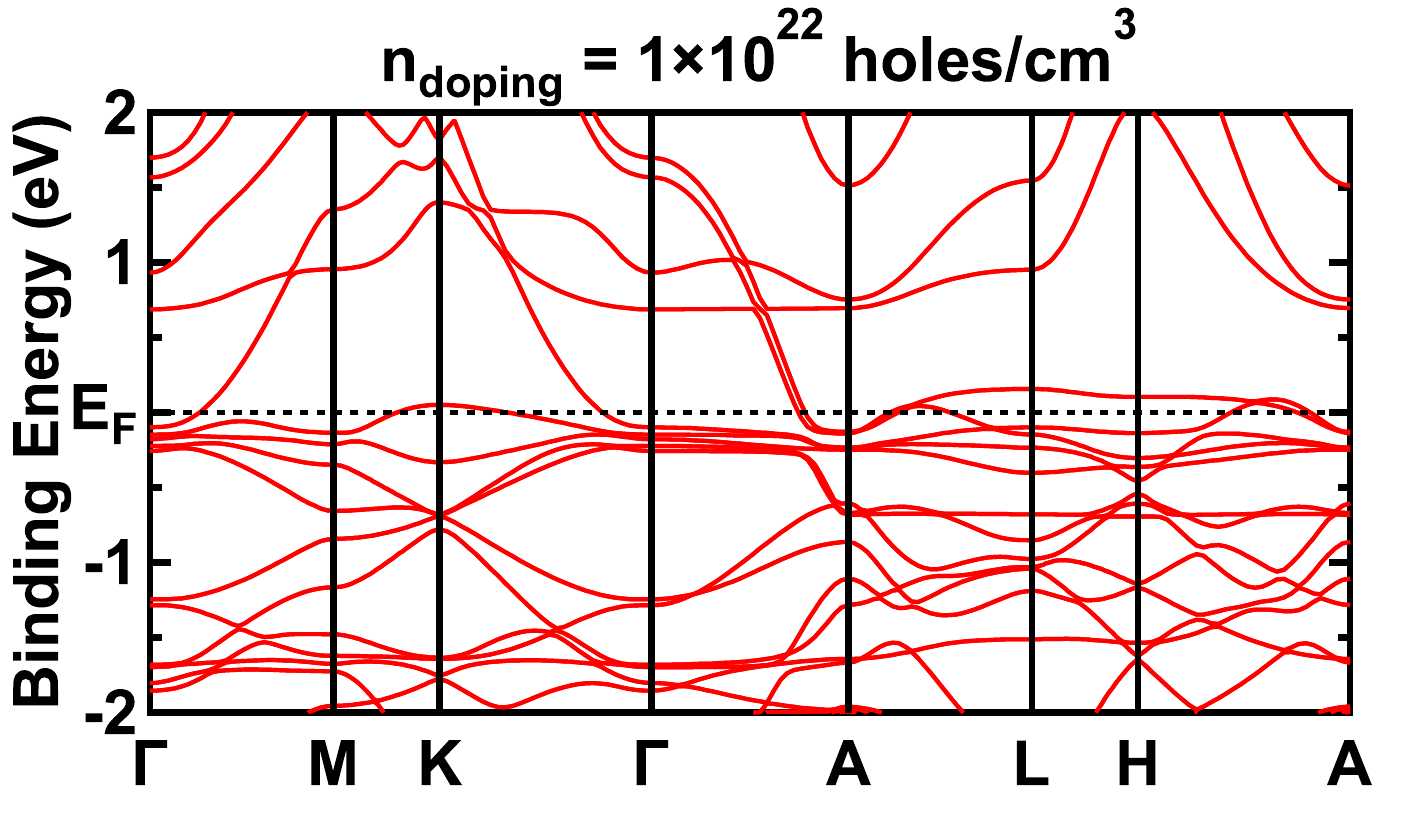}
    }
    \subfloat[\label{fig:DFT_0p5hole}]{
    \includegraphics[width=0.49\textwidth]{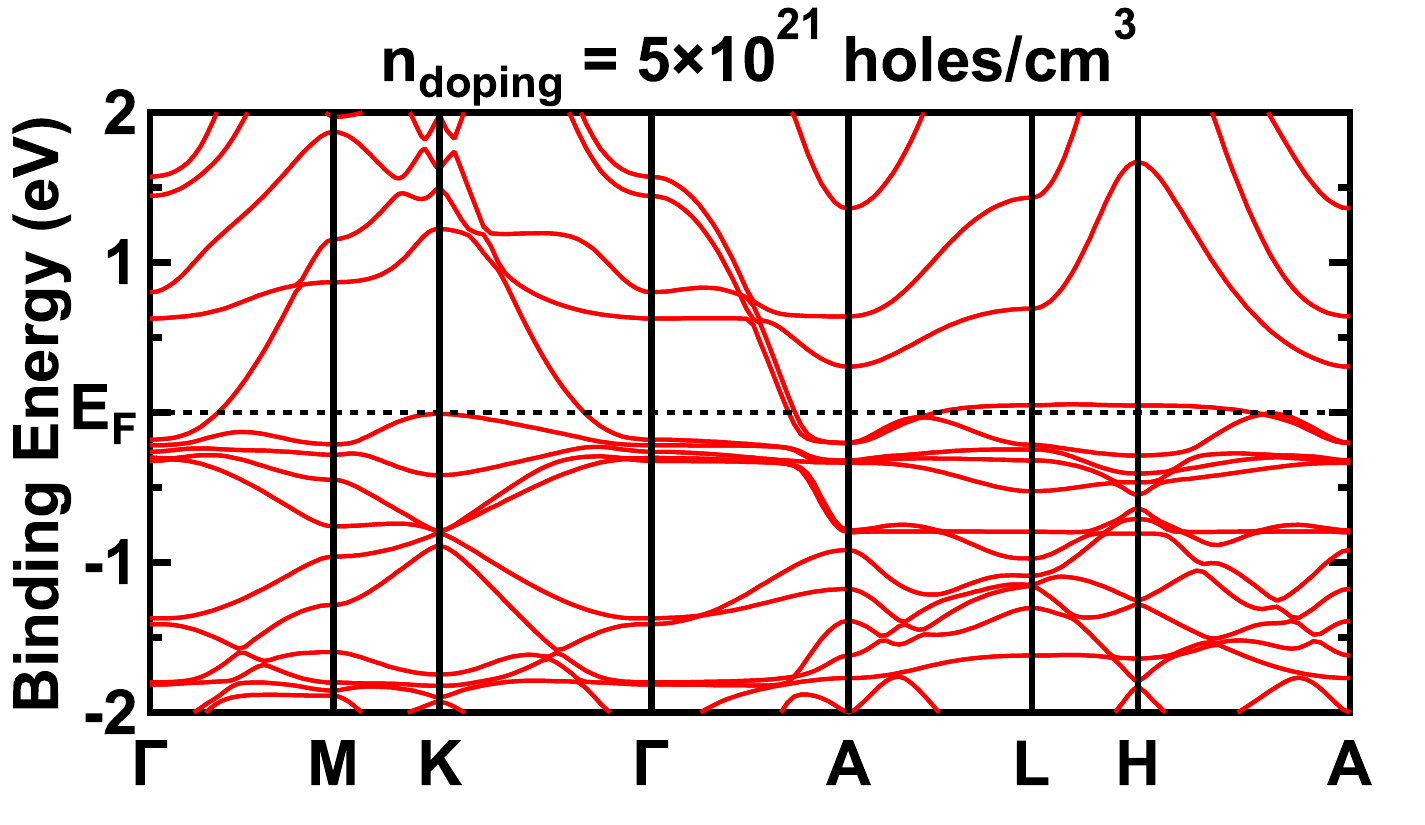}
    }\hfill
    \subfloat[\label{fig:DFT_pristine}]{
    \includegraphics[width=0.49\textwidth]{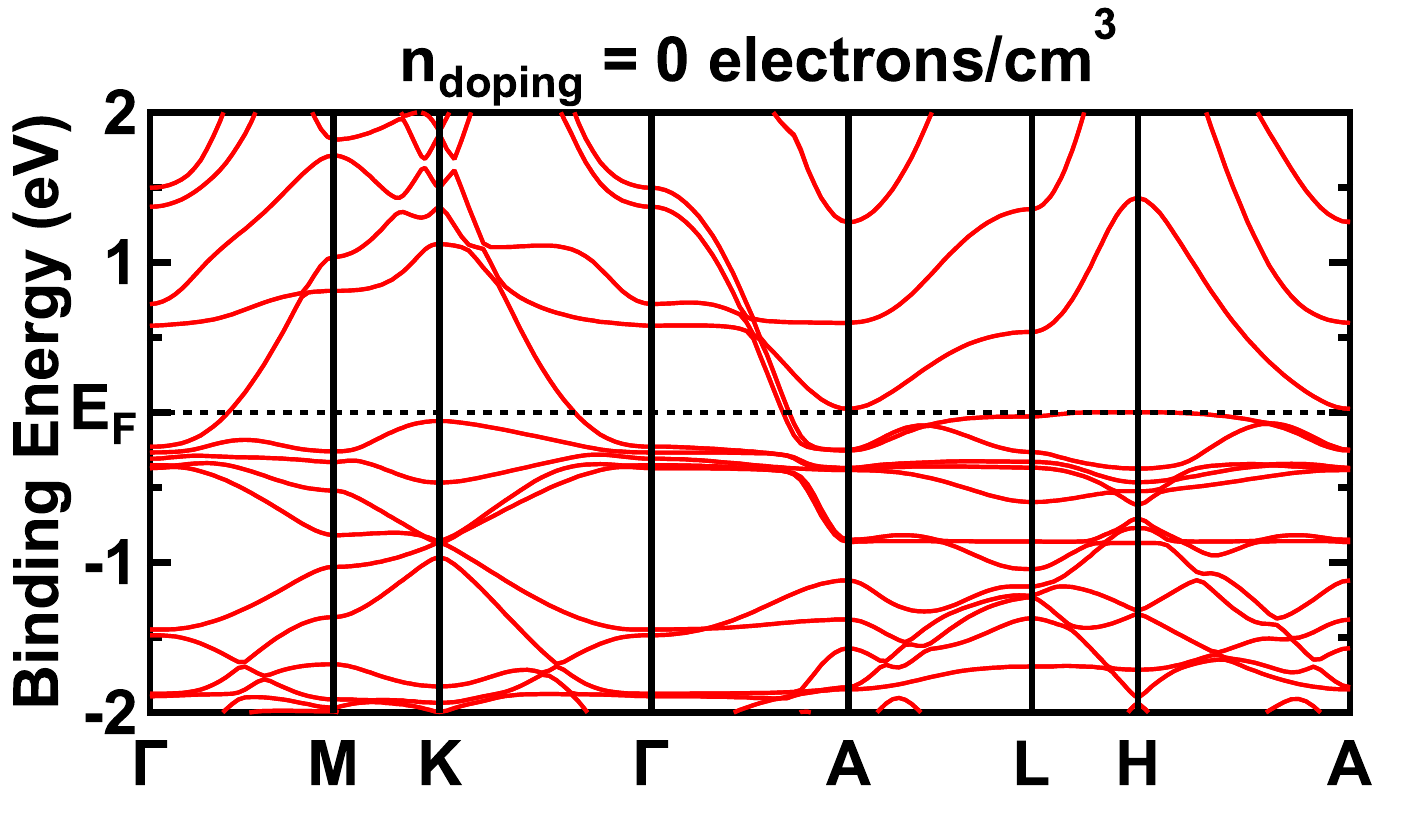}
    }
    \subfloat[\label{fig:DFT_0p5el}]{
       \includegraphics[width=0.49\textwidth]{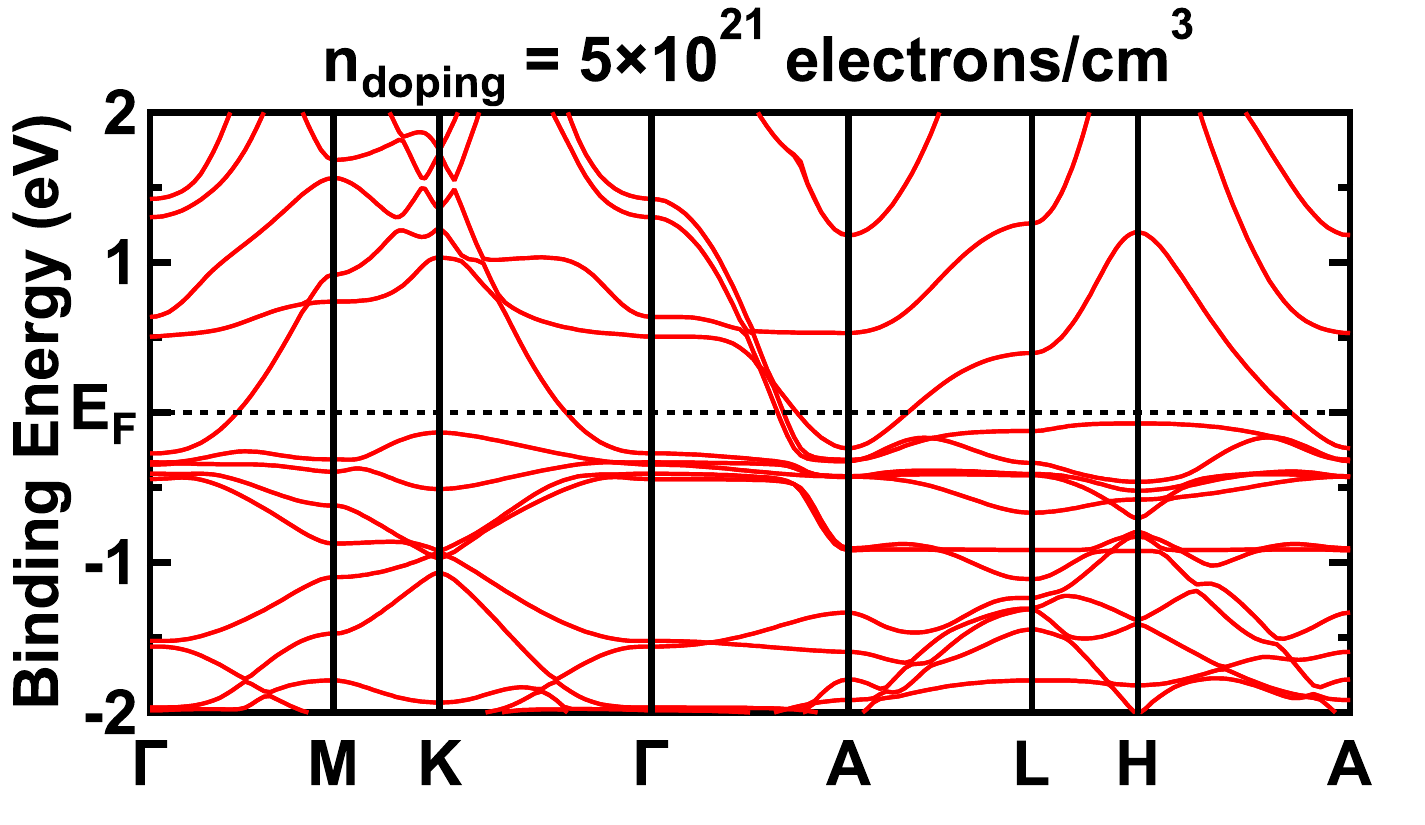}
    }\hfill    
    \subfloat[\label{fig:DFT_1el}]{
    \includegraphics[width=0.49\textwidth]{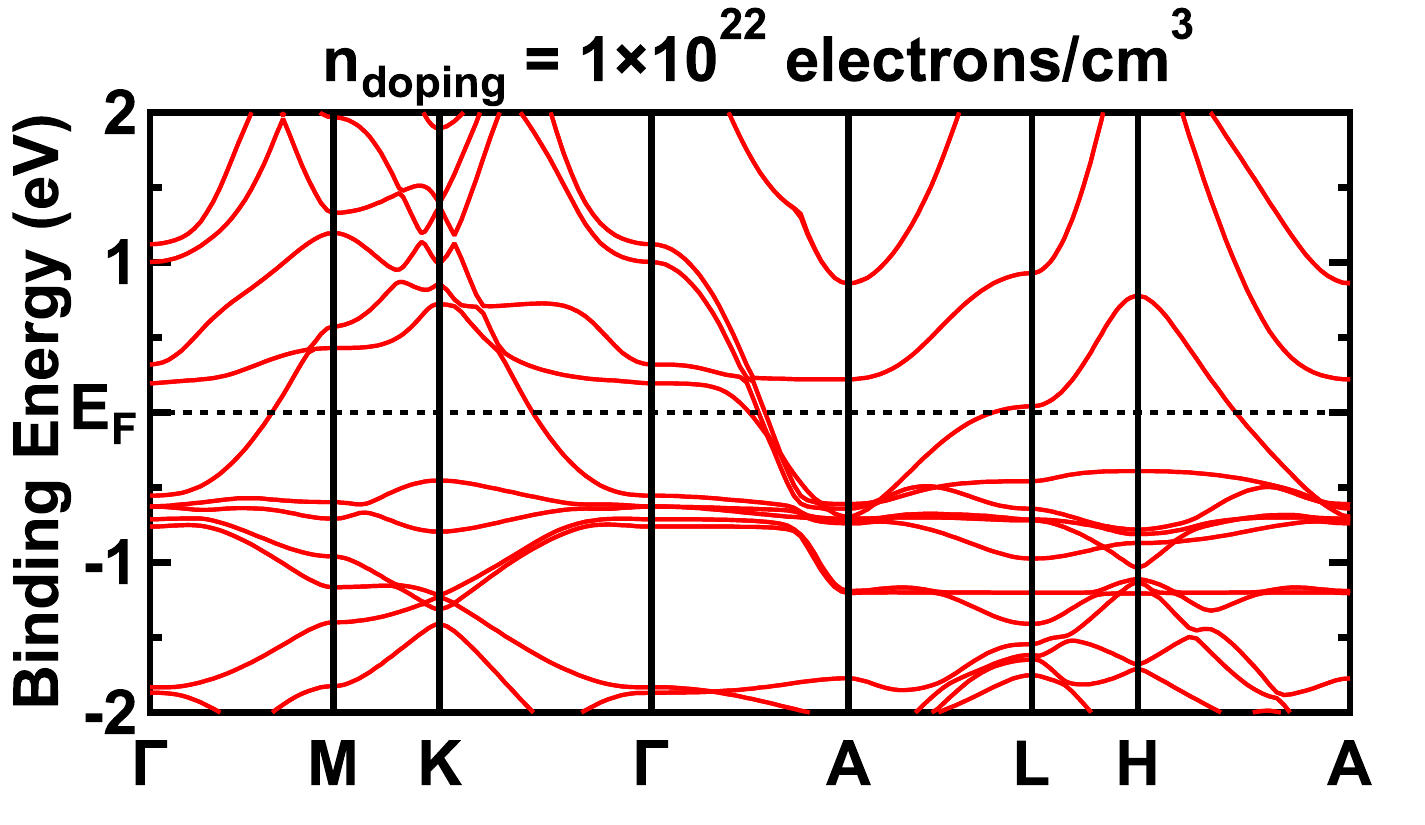}
    }\hfill
    \caption{\label{fig:DFT_Supp} Additional DFT calculation results with doping levels of (a) 1.0$\times$10$^{22}$\,holes/cm$^3$, (b) 1.0$\times$10$^{22}$\,holes/cm$^3$, (c) 0, (d) 0.5$\times$10$^{22}$\,electrons/cm$^3$, (e) 1.0$\times$10$^{22}$\,electrons/cm$^3$, respectively.
    } 
\end{figure}

We performed additional DFT calculations with several representative doping levels varying from 1.0$\times$10$^{22}$\,holes/cm$^3$ to 1.0$\times$10$^{22}$\,electrons/cm$^3$, as shown in Figure~\ref{fig:DFT_Supp}.

\newpage

\section{Estimation of Longitudinal Conductivity and Hall Conductivity}

The calculation of transport properties was performed by numerical integration on a three-dimensional grid with 0.01\,\AA$^{-1}$ step size.
By assuming an isotropic relaxation time $\tau_{i}$ for each band, we calculated $\displaystyle\frac{\sigma_{xx}^{(i)}}{\tau_{i}}$, $\displaystyle\frac{\sigma_{zz}^{(i)}}{\tau_{i}}$, and $\displaystyle\frac{\sigma_{H}^{(i)}}{\tau_{i}^2}$, which only depend on the band dispersion $\displaystyle\varepsilon^{(i)}(\textbf{k})$:
\begin{equation}\label{eq:equationS1}
\frac{\sigma_{xx}^{(i)}}{\tau_{i}}=-\frac{e^2}{4\pi^3\hbar^2}\int\left(\frac{\partial\varepsilon}{\partial k_{x}}\right)^{2}\frac{\partial f}{\partial \varepsilon}d\textbf{k}
\end{equation}
\begin{equation}\label{eq:equationS2}
\frac{\sigma_{zz}^{(i)}}{\tau_{i}}=-\frac{e^2}{4\pi^3\hbar^2}\int\left(\frac{\partial\varepsilon}{\partial k_{z}}\right)^{2}\frac{\partial f}{\partial \varepsilon}d\textbf{k}
\end{equation}
\begin{equation}\label{eq:equationS2}
\frac{\sigma_{H}^{(i)}}{\tau_{i}^2}=-\frac{e^3}{4\pi^3\hbar^4}\int\left[\left(\frac{\partial\varepsilon}{\partial k_{x}}\right)^{2} \left(\frac{\partial^{2}\varepsilon}{\partial k_{y}^{2}}\right) -\left(\frac{\partial\varepsilon}{\partial k_{x}}\right)\left(\frac{\partial\varepsilon}{\partial k_{y}}\right)\left(\frac{\partial^{2}\varepsilon}{\partial k_{x}\partial k_{y}}\right)\right]\frac{\partial f}{\partial \varepsilon}d\textbf{k}
\end{equation}
The results are summarized in Table~\ref{tab:tableS2}.
Taking typical relaxation times of 10$^{0}\sim$10$^{1}$\,fs in metals~\cite{palenskis2018phonon}, the in-plane longitudinal conductivity $\sigma_{xx}$ is on the order of 10$^{5}\sim$10$^{6}$\,S/m, which agrees with the experimental result of 9.5$\times$10$^{5}$\,S/m.
Furthermore, the calculation results suggest that the out-of-plane longitudinal conductivity $\sigma_{zz}$ is likely to be more than 1 order of magnitude larger than the in-plane conductivity $\sigma_{xx}$ for bands II, III, and IV.
This is consistent with the observed large anisotropic conductivity in bulk CoSn~\cite{huang2022flat}.

\begin{table*}[h]
\begin{tabular*}{\textwidth}{c @{\extracolsep{\fill}} ccccc}
 Band index                     & I    & II    & III (FB1)   & IV  \\\hline
 $\sigma_{xx}$/$\tau$ (10$^3$\,/($\Omega\cdot$m$\cdot$fs))        & 122 & 1.3  & 0.8  & 9.4   \\
 $\sigma_{zz}$/$\tau$ (10$^3$\,/($\Omega\cdot$m$\cdot$fs))        & 849 & 1035  & 537 & 11.3    \\
 $\sigma_{H}$/$\tau^2$ (m/(($\Omega^{2}\cdot$C$\cdot$fs$^{2}$)))    & -16  & 0.7 & -0.02  & -2.4      \\
\end{tabular*}
\caption{\label{tab:tableS2} Calculated transport properties from band structures.}
\end{table*}

\newpage

\bibliography{CoSn.bib}